\documentclass[aps,pre,10pt,onecolumn,superscriptaddress,balancelastpage,showpacs,letterpaper,lengthcheck]{revtex4-2}
\usepackage{centernot}
\usepackage{graphicx}
\usepackage{amsmath}
\usepackage{times}
\usepackage{amssymb}
\usepackage{mathrsfs}
\usepackage{chemarr}
\usepackage{xcolor}
\usepackage{url}
\usepackage{version}
\definecolor{linkcolor}{rgb}{0,0,0.6}
\usepackage[pdftex,colorlinks=true,
pdfstartview=FitV,
linkcolor= linkcolor,
citecolor= linkcolor,
urlcolor= linkcolor,
hyperindex=true,
hyperfigures=false]
{hyperref}

\newcommand{\mD}{\mathbb{D}}

\newcommand{\mH}{\mathbb{H}}

\newcommand{\M}{\mathbb{M}}
\newcommand{\Mm}{\mathbb{M}^{-1}}
\newcommand{\mM}{\M}

\newcommand{\cA}{\mathcal{A}}
\newcommand{\cD}{\mathcal{D}}

\newcommand{\cK}{\mathcal{K}}
\newcommand{\cL}{\mathcal{L}}
\newcommand{\cP}{\mathcal{P}}

\newcommand{\cS}{\mathcal{S}}
\newcommand{\cO}{\mathcal{O}}

\newcommand{\bet}{\boldsymbol{\eta}}
\newcommand{\dd}{\text{d}}
\newcommand{\ee}{\text{e}}
\newcommand{\bff}{\textbf{f}}
\newcommand{\bg}{\textbf{g}}
\newcommand{\bu}{\textbf{u}}

\newcommand{\p}{\partial}
\newcommand{\bp}{\textbf{p}}
\newcommand{\bq}{\textbf{q}}
\newcommand{\br}{\textbf{r}}
\newcommand{\bv}{\textbf{v}}
\newcommand{\bx}{\textbf{x}}
\newcommand{\by}{\textbf{y}}

\newcommand{\R}{\text{\tiny R}}

\newcommand{\eff}{\text{eff}}
\newcommand{\eps}{\varepsilon}

\newcommand{\sS}{\text{\tiny S}}

\newcommand{\Vr}{(V* \rho)}

\newcommand{\avg}[1]{\left \langle #1 \right \rangle}

\newcommand{\avgg}[1]{\big \langle #1 \big \rangle}
\newcommand{\pnt}[1]{\left ( #1 \right)}
\newcommand{\pntg}[1]{\big ( #1 \big)}
\newcommand{\brt}[1]{\left [ #1 \right]}
\newcommand{\brtg}[1]{\big [ #1 \big]}
\newcommand{\abs}[1]{\left | #1 \right|}

\newcommand{\cur}[1]{\left \{ #1 \right \}}

\newcommand{\f}[2]{\frac{ #1}{#2}}

\newcommand\bbar[1] { {\overline{\overline{#1}}}}
\newcommand\bbbar[1] { \overline{\overline{\overline{#1}}}}

\usepackage{tikz}
\begin{document}

\title{Statistical Mechanics of Active Ornstein Uhlenbeck Particles}

\author{David Martin}
\affiliation{Universit\'e de Paris, Laboratoire Mati\`ere et Syst\`emes Complexes (MSC), UMR 7057 CNRS,F-75205 Paris,  France}

\author{J\'er\'emy O'Byrne}
\affiliation{Universit\'e de Paris, Laboratoire Mati\`ere et Syst\`emes Complexes (MSC), UMR 7057 CNRS,F-75205 Paris,  France}

\author{Michael E. Cates}
\affiliation{DAMTP, Centre for Mathematical Sciences, University of Cambridge, Wilberforce Road, Cambridge CB3 0WA, UK}

\author{\'Etienne Fodor}
\affiliation{DAMTP, Centre for Mathematical Sciences, University of Cambridge, Wilberforce Road, Cambridge CB3 0WA, UK}

\affiliation{Department of Physics and Materials Science, University of Luxembourg, L-1511 Luxembourg}

\author{Cesare Nardini}
\affiliation{Service de Physique de l'\'Etat Condens\'e, CNRS UMR 3680, CEA-Saclay, 91191 Gif-sur-Yvette, France}
\affiliation{DAMTP, Centre for Mathematical Sciences, University of Cambridge, Wilberforce Road, Cambridge CB3 0WA, UK}

\author{Julien Tailleur}
\affiliation{Universit\'e de Paris, Laboratoire Mati\`ere et Syst\`emes Complexes (MSC), UMR 7057 CNRS,F-75205 Paris,  France}

\author{Fr\'ed\'eric van Wijland}
\affiliation{Universit\'e de Paris, Laboratoire Mati\`ere et Syst\`emes Complexes (MSC), UMR 7057 CNRS,F-75205 Paris,  France}

\date{\today}

\begin{abstract}
We study the statistical properties of Active Ornstein Uhlenbeck
particles (AOUPs). In this simplest of models, the Gaussian white
noise of overdamped Brownian colloids is replaced by a Gaussian
colored noise. This suffices to grant this system the hallmark
properties of active matter, while still allowing for analytical
progress. We study in detail the steady-state distribution of AOUPs in
the small persistence time limit and for spatially varying activity.
At the collective level, we show AOUPs to experience motility-induced
phase separation both in the presence of pairwise forces or due to
quorum-sensing interactions. We characterize both the instability
mechanism leading to phase separation and the resulting phase
coexistence. We probe how, in the stationary state, AOUPs depart from
their thermal equilibrium limit by investigating the emergence of
ratchet currents and entropy production. In the small persistence time
limit, we show how fluctuation-dissipation relations are
recovered. Finally, we discuss how the emerging properties of AOUPs
can be characterized from the dynamics of their collective modes.
\end{abstract}


\maketitle


\section{Introduction}

The hallmark of activity is the lack of connection at the microscopic
scale between the injection of energy and its dissipation. While the
latter is provided by the environment which, like for an equilibrium
colloidal particle, can be an equilibrated liquid, the former arises
from the consumption of an independent energy source, whose origin
varies from systems to systems. In biology, energy injection typically
results from the hydrolysis of ATP, which drives almost all biological
processes such as cell motion~\cite{SawNature2017,KawaguchiNature2017}
or intracellular trafficking~\cite{guo2014probing,Ahmed2018}.
Synthetic active particles, on the other hand, have been powered using
a large range of external sources, whose natures range from
chemical~\cite{howse2007self,palacci2010sedimentation,volpe2011microswimmers}
to
electrical~\cite{Bricard:2013:Nature,nishiguchi2015mesoscopic,van2019interrupted,yan2016reconfiguring}
to
mechanical~\cite{volfson2004anisotropy,narayan2007long,deseigne2010collective}.

At a theoretical level, the modelling of active systems has been
equally diverse, starting either at a coarse-grained, phenomenological
level~\cite{toner1995long,kruse2004asters,marchetti2013hydrodynamics,wittkowski2014scalar}
or directly at the microscopic scale, using various models of
self-propelled
particles~\cite{Vicsek1995PRL,cates2012diffusive,romanczuk2012active,Fodor2018}. In
the latter case, the standard dynamics of $N$ self-propelled particles
typically writes, ignoring inertial effects, as
\begin{equation}\label{eq:dyn}
   \dot \br_i = - \mu \nabla_i \Phi + \bv_i ,
\end{equation}
where $\br_i$ describes the position of particle $i$, $\bv_i$ its
self-propulsion velocity, $\mu$ the particle mobility, and we have
included a putative interaction potential
$\Phi(\br_1,\dots,\br_N)$. The type of active particles is then
specified by the properties of their self-propulsions $\bv_i$. The
most studied examples are probably run-and-tumble particles
(RTPs)~\cite{Schnitzer:93} and active Brownian particles
(ABPs)~\cite{Fily:12}, for which $\bv_i$ has a constant modulus and
reorient either stochastically at a given rate for RTPs or following
rotational diffusion for ABPs. These microscopic models and their
variants have led to a wealth of surprising behaviours, among which
the accumulation at hard boundaries~\cite{Gompper:09,Cates:09},
the generation of currents by asymmetric
obstacles~\cite{galajda2007wall,Cates:09,di2010bacterial,sokolov2010swimming,Koumakis2014,Pietzonka2019},
the emergence of collective
motion~\cite{Vicsek1995PRL,Gregoire:2004:PRL}, and motility-induced
phase separation (MIPS)~\cite{Tailleur:08,Fily:12,Cates2015MIPS}.

Whatever the microscopic non-equilibrium origin of self-propulsion,
these rich behaviors would be impossible if the $\bv_i$'s were well
described by Gaussian white noises, which would make
Eq.~\eqref{eq:dyn} an equilibrium dynamics. ABPs and RTPs differ from
this simpler case by both the non-Gaussian nature and the persistence
of their self-propulsion. Over the past few years, a desire to
pinpoint the fundamentals of active matter has led to study simplified
models, in which only the non-Gaussian nature~\cite{fodor2018non} or
the
persistence~\cite{Szamel:14,Maggi:15,Marconi:15,Szamel:15,Brader:15,Fodor2016PRL,Wittmann_2017_1,Wittmann_2017_2,marini2017pressure,caprini2019entropy,caprini2019activity}
is retained. In this article, we consider the latter case, referred to
as Active Ornstein-Uhlenbeck particles (AOUPs), which, besides their
theoretical motivations, have also been used to model the motion of
passive tracers in an active bath of bacteria~\cite{Maggi:14a,
  Koumakis2014} as well as for the collective dynamics of
cells~\cite{deforet2014emergence,hakim2017collective}.

The self-propulsion velocities of AOUPs are given by $N$ independent
Ornstein-Uhlenbeck processes:
\begin{equation}\label{eq:OU}
  \tau \dot \bv_i = - \bv_i + (2D)^{1/2} \bet_i ,
\end{equation}
where $\{ \bet_i \}$ forms a set of zero-mean Gaussian white noises
with correlations $ \avg{ \eta_{i\alpha} (t) \eta_{j\beta} (t')} =
\delta_{ij} \delta_{\alpha\beta} \delta(t-t')$, with Greek indices
running over spatial dimensions. It follows that $ \cur{ \bv_i } $
forms a set of zero-mean colored Gaussian noises with correlations
\begin{equation}\label{eq:v}
  \avg{ v_{i\alpha} (t) v_{j\beta} (0) } = \delta_{ij} \delta_{\alpha\beta} \f{D}{\tau} \ee^{ - \abs{t} / \tau } ,
\end{equation}
where $D$ controls the noise amplitude, and $\tau$ quantifies the
noise persistence. In the limit of vanishing $\tau$, the
self-propulsion velocities $\bv_i$ reduce to Gaussian white noises of
correlation
$ \avg{ v_{i\alpha} (t) v_{j\beta} (s) } = \delta_{ij}
\delta_{\alpha\beta} 2 D \delta(t-s)$. For finite $\tau$, on the
contrary, the temporal correlations of the `active noise' $\bv_i$ are
not matched by a corresponding memory kernel for the instantaneous
damping $\gamma\equiv\mu^{-1}$, so that Eq.~\eqref{eq:dyn} does not
satisfy a FDT of the second kind: it describes a non-thermal
process~\cite{berthier2013non}. The correlation time $\tau$ hence
controls the nonequilibrium properties of the dynamics.

In this article, we study AOUPs both at single-particle and collective
levels, and we characterize their static properties as well as their
non-equilibrium dynamical features. In section~\ref{sec:fpe}, we first
discuss various limiting cases in which the steady-state distributions
of AOUPs can be computed analytically. We discuss in particular the
small-persistence-time regime, as well as the impact of non-uniform
persistence and activity on the spatial distribution of particles. In
section~\ref{sec:MIPS}, we show that AOUPs can undergo MIPS, both in
the presence of pairwise purely repulsive forces and due to
quorum-sensing interactions if the typical velocity, $\sqrt{D/\tau}$,
is a sufficiently strongly decreasing function of the local particle
density. In section~\ref{sec:TRS}, we then turn to characterize
dynamically the departure from equilibrium of AOUPs. We first consider
the occurence of particle currents in the presence of ratchet
potentials and we then fully characterize the breakdown of
time-reversal symmetry. At linear order in $\tau$, we show AOUPs to
admit an effective, non-Boltzmann equilibrium regime, which we
characterize dynamically in section~\ref{sec:fdr} by deriving the
corresponding generalized fluctuation-dissipation theorem (FDT).
Finally, we discuss in section~\ref{sec:coll_modes} the explicit
coarse-graining of $N$ interacting AOUPs by constructing the dynamics
of their collective modes. Note that
sections~\ref{sec:smalltau1},~\ref{sec:sig},
and~\ref{subsec:perturbing-potential} correspond to detailed
presentations of results announced in~\cite{Fodor2016PRL} while
section~\ref{subsec:Fox} is a short review on the application of UCNA
and Fox theory to AOUPs. All other sections correspond to new
material.

\section{Steady state distribution}\label{sec:fpe}

This section is devoted to the analytical characterization of the
stationary distribution of AOUPs. We begin in~\ref{sec:smalltau1} with
the short-persistent-time regime, which has attracted a lot of
interest
recently~\cite{Maggi:15,Marconi:15,Brader:15,Fodor2016PRL,Wittmann_2017_1,Wittmann_2017_2,marini2017pressure,caprini2019entropy,caprini2019activity,bonilla2019active}. We
present an expansion to second order in $\tau$ and show
in~\ref{subsec:SM-1particle-1d} how, for a single degree of freedom,
it can be extended to higher order. The gap with standard methods
developed to characterize the steady-state of particles powered by
Gaussian colored noises~\cite{Fox:86a, Fox:86b,Hanggi:87, Cao:93} is
then bridged in~\ref{subsec:Fox}. Finally, the steady-state of AOUPs
with spatially varying activity and persistence is discussed
in~\ref{sec:varying}. Note that
sections~\ref{sec:smalltau1}-\ref{subsec:Fox} are detailed derivations
of the results presented in~\cite{Fodor2016PRL}, especially for the steady-state distribution Eq.~\eqref{eq:Ps_eff}, and discussions of
these results with respect to the existing literature; their content is mostly
technical and they can be omitted during a first reading of the article.

\subsection{Systematic perturbative derivation}\label{sec:smalltau1}

Let us start by introducing the particle velocities $ \bp_i = \dot \br_i $
to recast the dynamics of $N$ interacting AOUPS into:
\begin{equation}\label{eq:dyn_active_g}
  \tau \dot \bp_i = - \bp_i - ( 1 + \tau \bp_j \cdot \nabla_j ) \nabla_i \Phi + ( 2 D )^{1/2} \bet_i ,
\end{equation}
where the mobility has been set to unity.  Setting $\tau=0$ in both
sides of Eq.~\eqref{eq:dyn_active_g} leads to the standard overdamped
equilibrium dynamics:
\begin{equation}\label{eq:K}
 \bp_i = - \nabla_i \Phi + (2D)^{1/2} \bet_i ,
\end{equation}
Note that setting $\tau=0$ only in the right-hand side of
Eq.~\eqref{eq:dyn_active_g} also leads to an equilibrium process: the
underdamped Kramers equation. More surprisingly, setting $\tau=0$ in
the left-hand side of Eq.~\eqref{eq:dyn_active_g} also leads to an
equilibrium dynamics, albeit with non-trivial mobility and effective
potential~\cite{Maggi:15}. This path will be detailed
in~\ref{subsec:Fox} and we now discuss how a small $\tau$ expansion
of~\eqref{eq:dyn_active_g} can instead be \textit{systematically}
derived.

Here we follow the treatment introduced in~\cite{Fodor2016PRL} which
works with the $\{\br_{i},\bp_{i}\}$ variables. In the spirit of~\cite{Hagan:87, Hanggi:87}, we first introduce the scaled variables $\bar t \equiv \tau^{-1/2} t$ and $\bar \bp_i \equiv \bp_i \tau^{1/2}$. As a result, the
stationary distribution satisfies $ \cL P_\sS ( \cur{ \br_i, \bar\bp_i } )
= 0 $, where the operator $\cL$ reads
\begin{equation}\label{eq:FP-operator-cesare}
	\begin{aligned}
		\cL &= - \bar\bp_{i} \cdot\nabla_i + \tau^{-1/2} \f{\p}{ \p \bar p_{i\alpha} } \pnt{ \bar p_{i\alpha} + \tau \bar p_{j\beta} \f{ \p^2 \Phi }{ \p r_{i\alpha} r_{j\beta} } }
		\\
		& \quad + \f{\p}{ \p \bar p_{i\alpha} } \f{ \p \Phi }{ \p r_{i\alpha} } + D \tau^{-1/2} \f{\p^2}{ \p \bar p_{i\alpha}^2 } .
	\end{aligned}
\end{equation}
Here, and in the following, summations over repeated indices are
implicit. (This includes terms like $\bar \bp_{i}^{2}$ or
$\f{\p^2}{ \p \bar p_{i\alpha}^2 }$.)  To compute the stationary
distribution, we propose the following ansatz:
\begin{align}\label{eq:AnsatzPs}
P_\sS (\br,\bar \bp) \sim e^{-\frac{\Phi}{D}-\frac{\bar \bp_{i}^{2}}{2D}}
\left(1+
\sum_{n=2}^{\infty} \tau^{\frac{n}{2}} A_n\pnt{\br, \bar \bp}\right)\,
\end{align}
where, for convenience, we define $A_0\equiv 1$ and $A_1\equiv 0$, and we introduce the notation $\br=\{\br_i\},\bar\bp=\{\bar\bp_i\}$, which lightens the notations in the many-particle case. Note that, for normalization purposes, $\int e^{-\frac{\Phi}{D}-\frac{\bar \bp_{i}^{2}}{2D}} A_n d\bar{{\bf p}}d{\bf r}$ has to vanish.
We then obtain a set of recursive equations for the $A_n$ equating every order in $ \tau^{1/2} $
\begin{align}
\label{Aeq}
&\left(\bar p_{i\alpha}\frac{\partial}{\partial \bar p_{i\alpha}}-D\frac{\partial^{2}}{\partial \bar p_{i\alpha}^{2}}\right)A_{n}
= f_n \pnt{ \br, \bar \bp}
\end{align}
where
\begin{align}
&f_n \pnt{ \br, \bar \bp }
=-\bar p_{i\alpha}\frac{\p A_{n-1}}{\p r_{i\alpha}}+\frac{\partial \Phi}{\partial r_{i\alpha}}\frac{\partial A_{n-1}}{\partial \bar p_{i\alpha}}
\nonumber + \frac{\partial ^{2}\Phi}{\partial r_{i\alpha}^{2}}A_{n-2}\\
&-\frac{\bar p_{i\alpha}\bar p_{j\beta}}{D}\frac{\partial \Phi}{\partial r_{i\alpha}\partial r_{j\beta}}A_{n-2}
+\bar p_{i\alpha}\frac{\partial^{2}\Phi}{\partial r_{i\alpha} \partial r_{j\beta}}\frac{\partial A_{n-2}}{\partial \bar p_{j\beta}}\,.
\end{align}

Inspection of~(\ref{Aeq}) suggests an ansatz for the $A_n$ in the form
of degree-$n$ polynomials in the momenta. (If this is assumed for
$A_k$ with $k\leq n$, then $A_{n+1}$ is a polynomial of degree
$(n+1)$.) It can be also checked that $A_{2n}$ contains only even
terms in the momenta and $A_{2n+1}$ only odd ones. This results from
the symmetry of the equation $ \cL P_\sS = 0 $ under the
transformation $ \cur{ \tau^{1/2}, \bar \bp } \to - \cur{ \tau^{1/2},
  \bar \bp } $. We use the following notation
 \begin{align}
\label{GeneralizedPolynoms}
A_{n}&=
\bar p_{i_{1},\alpha_1}...\bar p_{i_{n},\alpha_n}\frac{a^{(n,n)}_{i_{1},...,i_{n}, \alpha_1,...,\alpha_n}}{n!} \\
\nonumber &
+\bar p_{i_{1},\alpha_1}...\bar p_{i_{n-2},\alpha_{n-2}}\frac{a^{(n,n-2)}_{i_{1},...,i_{n-2},\alpha_{1},...,\alpha_{n-2}}}{(n-2)!}+...
\end{align}
where the $a^{(m,n)}$'s depend on the particles' positions. Note that
$A_n$ contains a $\bar \bp$-independent term $a^{(n,0)}$ only if $n$ is
even. Note also that~\eqref{GeneralizedPolynoms} is a local function of the momenta, which could be restrictive, but still allow for non-local dependence on the particle positions, something known to be important for active particles~\cite{van1984activation,Solon:15b}.

Plugging the expression (\ref{GeneralizedPolynoms}) for $A_n$ in
(\ref{Aeq}), and equating order by order in $\bar \bp$, yields explicit
expressions of all tensors $a^{(n,m)}$ for $0<m\leq n$. For even $n$,
this leaves $a^{(n,0)}$ unconstrained (whereas $a^{(2k+1,0)}=0$ by
definition) but constrains $a^{(n-2,0)}$. For
instance, we get for $n=2,3,4$:
\begin{align}
A_{2}&=-\frac{(\bar \bp_i\cdot \nabla_i)^2 \Phi}{2D}+
a^{(2,0)}(\{\br_{i}\}) \\
A_{3}&=\frac{(\bar \bp_i\cdot\nabla_i)^3 \Phi}{6D}
+(\bar \bp_i\cdot\nabla_i)\nabla_j^2\Phi \nonumber \\
  &
 -\frac{1}{2D}(\bar \bp_{i}\cdot \nabla_i)(\nabla_j\Phi)^3
 -(\bar \bp_{i}\cdot \nabla_i) a^{(2,0)} \label{eq:Pss-A3}\\
A_4 &= \frac{1}{8D^2}\left[(\bar \bp_i\cdot\nabla_i)^2 \Phi\right]\left[(\bar \bp_j\cdot\nabla_j)^2 \Phi\right] - \frac{(\bar \bp_i\cdot\nabla_i)^4\Phi}{24D} \nonumber \\ &\frac{(\bar \bp_i\cdot\nabla_i)^2\left[a^{(2,0)}-\nabla^2\Phi\right]}{2}+\frac{3}{4D}\frac{\p \Phi}{\p r_{j,\alpha}}\frac{\p}{\p r_{j,\alpha}}(\bar \bp_i\cdot\nabla_i)^2\Phi  \nonumber \\
& +\frac{1}{2D}\frac{\p\left[(\bar \bp_j\cdot\nabla_j)\Phi\right]}{\p r_{i,\alpha}}\frac{\p\left[(\bar \bp_k\cdot\nabla_k)\Phi\right]}{\p r_{i,\alpha}}
+a^{(4,0)}
\end{align}
The lowest order in $\bar {\bf p}$ of Eq.~\eqref{Aeq} for $A_{n}$ of the form~\eqref{GeneralizedPolynoms} also yields
\begin{align}
\label{SolvabCond}
\left(D\frac{\partial^{2} }{\partial r_{i\alpha}^{2}}-\frac{\partial \Phi}{\partial r_{i\alpha}}\frac{\partial }{\partial r_{i\alpha}}\right)
a^{(n-2,0)}
=
g_{n}
\end{align}
where $g_{n}$ are functions of the particle positions which
can be computed explicitly. Finding a solution to (\ref{SolvabCond}) thus provides a closed expression for the expansion up to order $n-2$.

From the Fredholm alternative theorem, the condition under which
(\ref{SolvabCond}) admits a solution, and the expansion can be carried
out, is that $g_n$ is orthogonal to
$e^{-\Phi/D}$~\cite{fredholm1903classe}. This is always possible in
the small $D$ limit, following~\cite{nardiniperturbative2016}, but the
existence of solution to arbitrary order remains, in the general case,
an open problem. Here we show that it can be carried out explicitly up
to order $\tau^{3/2}$. $g_4$ indeed reads
\begin{align}
g_4=
&\frac{1}{D}\frac{\partial^{2}\Phi}{\partial r_{i\alpha}\partial r_{j\beta}}\frac{\partial\Phi}{\partial r_{i\alpha}}\frac{\partial\Phi}{\partial r_{j\beta}}
-\frac{5}{2}\frac{\partial^{3}\Phi}{\partial^{2} r_{i\alpha}\partial r_{j\beta}}\frac{\partial\Phi}{\partial r_{j\beta}} \\ \nonumber &
-\frac{\partial^{2} \Phi}{\partial r_{i\alpha}\partial r_{j\beta}}\frac{\partial^{2} \Phi}{\partial r_{i\alpha}\partial r_{j\beta}}
+\frac{3D}{2}\frac{\partial^{4}\Phi}{\partial^{2} r_{i\alpha}\partial^{2} r_{j\beta}}\,.
\end{align}
and (\ref{SolvabCond}) is solved, for $n=2$, by
\begin{align}
a^{(2,0)}&=
-\frac{1}{2D}\left(\nabla_i \Phi\right)^{2}
+\frac{3}{2}\nabla^2_i \Phi\,.
\end{align}
This yields the following expression for the stationary measure of $N$ interacting AOUPs, valid up to order $\tau^{3/2}$,
\begin{equation}\label{eq:Pss}
	\begin{aligned}
		P_\sS \sim & \, \ee^{ - \f{\Phi + \bar \bp_i^2 / 2 }{D} } \Big \{ 1 - \f{\tau}{2D} \brt{ \pnt{ \nabla_i \Phi }^2 + \pnt{ \bar \bp_i \cdot \nabla_i }^2 \Phi - 3 D \nabla_i^2 \Phi }
		\\
		& + \f{ \tau^{3/2} }{6 D} ( \bar \bp_i \cdot \nabla_i ) \brt{ (\bar \bp_j \cdot \nabla_j)^2 - 3 D \nabla_j^2 } \Phi + \cO \pnt{ \tau^2 } \Big \} .
	\end{aligned}
\end{equation}
The velocity distribution obtained by integrating over the particle position is Gaussian to first order in $\tau$, with the same variance as in~\cite{marconi2016velocity}, obtained within the UCNA framework that we discuss in Section \ref{subsec:Fox}. However, the $\tau^{3/2}$ order shows that the velocity distribution is non-Gaussian to this order.

Integrating over velocities, we obtain the many-body marginal
distribution, in position space, $ P_\sS ( \cur{ \br_i } ) $:
\begin{equation}\label{eq:Ps_eff}
	P_\sS ( \cur{ \br_i } ) \sim \exp \brt{ - \f{\Phi}{D} - \f{\tau}{2D} ( \nabla_i \Phi )^2 + \tau \nabla^2_i  \Phi + \cO \pnt{ \tau^2 } } .
\end{equation}
We can now use expression \eqref{eq:Ps_eff} to define an effective potential through $\Phi_\eff = - D \ln P_\sS ( \cur{ \br_i } )$ which provides an intuitive picture of how self-propulsion affects the bare steady-state $\exp(-\Phi/D)$.
The correction term $ \tau (\nabla_i \Phi)^2 / 2$ is always positive, irrespective of the repulsive or attractive nature of the force $-\nabla_i \Phi$: it drives particles away from regions of large forces. The second correction term $ - D \tau \nabla_i^2 \Phi $ is dominant for large values of $D$ and can take either sign. It favors the presence of particles in convex potential landscapes.
When the bare potential describes pairwise interactions, the corresponding effective potential contains three-body interactions stemming from the term $ \tau (\nabla_i \Phi)^2 / 2 $. Moreover, when $\Phi$ corresponds to repulsive interactions only, the associated $\Phi_\eff$ combines repulsive and attractive interactions~\cite{Brader:15,Marconi:15}. This captures how self-propulsion produces attractive effects out of repulsive forces, consistently with the enhanced tendency towards clustering reported experimentally for various colloidal systems~\cite{theurkauff2012dynamic,Palacci:13, Speck:13,io2017experimental,geyer2019freezing}.

It is interesting to note that similar functional forms to~\eqref{eq:Pss} and~\eqref{eq:Ps_eff} are encountered in many contexts, from the semi-classical expansion of the Boltzmann distribution in powers of $\hbar$~\cite{landauV-33} to the Hermitian form of the Fokker-Planck operator~\cite{VanKampenBook}. It would be interesting to know whether this is just a coincidence or reflects the presence of a deeper connection.

\subsection{One particle in one dimension}\label{subsec:SM-1particle-1d}
The systematic perturbative expansion presented in Sec. \ref{sec:fpe} can be carried out to any order in $\tau$
in the case of a single AOUP particle in $d=1$ for an arbitrary smooth external potential $\Phi$ in an infinite space.
In this case the solvability conditions (\ref{SolvabCond})  admits a solution to arbitrary order $n$, explicitly given by
\begin{equation}\label{eq:solvability-1d}
a^{(n-2,0)}(r) = \frac{1}{Z}
\int_0^{r} dy\,e^{-\frac{\Phi(y)}{D}}
\int_0^y dz\,e^{\frac{\Phi(z)}{D}} g_n(z)
\end{equation}
where $Z$ is a constant to be fixed by the normalization of the
stationary measure. In this case, the solution can be obtained
iteratively to arbitrary order, at the cost of expressions of
increasing complexity. The result of this procedure is illustrated in
Appendix~\ref{app:SM} where we give the full stationary distribution
$P_\sS ( r,p$ ) up to order $\tau^2$. Once the velocity is
integrated out, we find:
\begin{equation}
\label{eq:OnePartOneD}
\begin{aligned}
& P_\sS ( r )\sim  \exp\left[-\frac{\Phi}{D}+\tau\left(\Phi''(r)-\frac{\Phi'(r)^2}{2D}\right)+\tau^{2}\left(\frac{D\Phi^{(4)}(r)}{2} \right. \right.\\ \nonumber &\left.\left.+\frac{\int^{r} \Phi'(y)^2 \Phi^{(3)}(y) dy}{2D}-\Phi'(r)\Phi^{(3)}(r)-\frac{\Phi''(r)^{2}}{4}\right)+{\cal O}(\tau^{3})\right]\,,
\end{aligned}
\end{equation}
where $\Phi^{(n)}$ refers to the $n^{\rm th}$-order derivative of $\Phi$.
Interestingly, this result is compatible with a recent,
instanton-based derivation of the steady-state of AOUPs obtained
sending $D\to 0$ \textit{before} taking the $\tau\to 0$
limit~\cite{woillez2019nonlocal}. (Naturally, in this limit, only the
terms proportional to $D^{-1}$ in Eq.~\eqref{eq:OnePartOneD} survive.)

The case with periodic boundary condition is discussed in
Appendix~\ref{app:SM}.  Furthermore, the perturbation expansion can
also be generalized to the case of one particle in a central potential
in arbitrary dimension, as well as to the case of two interacting
particles with central forces. (We leave these cases to future works.)

Let us take advantage of working in a simpler, one-dimensional context
to address more subtle questions regarding the nature of our small
$\sqrt{\tau}$ expansion~\eqref{eq:AnsatzPs}. It is for instance
natural to ask whether this series admits a finite convergence radius
in $\tau$. Equation~\eqref{eq:AnsatzPs} indeed implicitly assumes that
$P_\sS$ is analytic in $\sqrt{\tau}$ which need not necessarily hold
for any potential.

\begin{figure}
  \def\Y{.9}
  \def\X{1.7}
  \begin{tikzpicture}

    \path (0,0)  node {\includegraphics[width=.45\columnwidth]{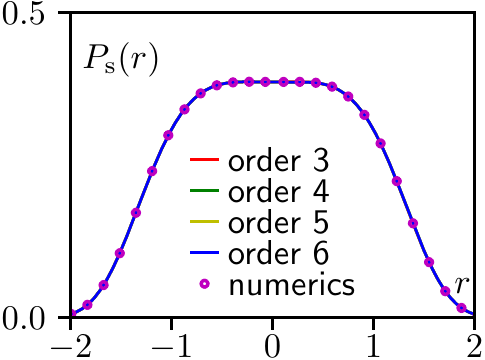}}
     (\X,\Y) node {a)};
    \path (4,0)  node {\includegraphics[width=.45\columnwidth]{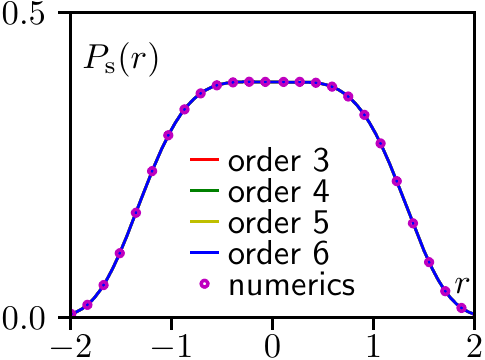}}
    (\X+4,\Y) node {b)};
    \path (0,-3)  node {\includegraphics[width=.45\columnwidth]{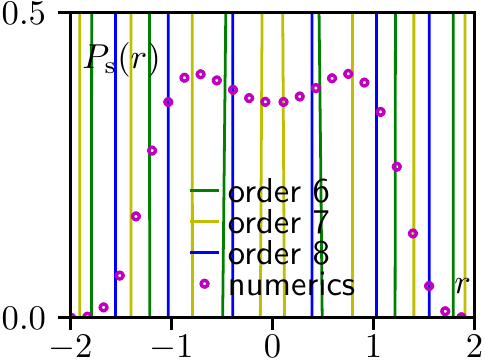}}
    (\X,\Y-3) node {c)};
    \path (4,-3)  node {\includegraphics[width=.45\columnwidth]{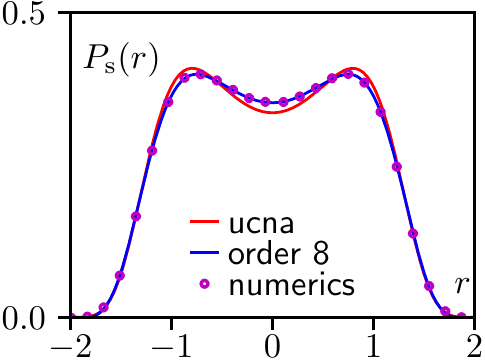}}
    (\X+4,\Y-3) node {d)};;

  \end{tikzpicture}
  \caption{Steady-state distribution of an AOUP in a confining potential $\Phi(x)=x^4/4$. {\bf Top:} For $\tau=0.01$, the series~\eqref{eq:seriesmarginal} converges rapidly and coincide with the numerics. The latter are obtained by integrating Eqs.~\eqref{eq:dyn} and~\eqref{eq:OU} using the Heun algorithm (See~\cite{martin2020aoup} for detail). ({\bf a}). The truncated Borel also describes correctly the data ({\bf b}). {\bf Bottom} For $\tau=0.2$, the series~\eqref{eq:seriesmarginal} is rapidly diverging and very far from the numerics ({\bf c}). On the contrary, the truncated Borel sum describes very well the data ({\bf d}).   }\label{fig:pade}
\end{figure}

To address this question, we consider the potential $\Phi(x)=x^4/4$,
at fixed $D$ and for two different values of $\tau$, and consider the
marginal in space of the full distribution:
\begin{equation}\label{eq:seriesmarginal}
  P_\sS(r)=\sum_n \bar A_{n}(r) \tau^n
\end{equation}
where
$\bar A_{n}(r)=\int dv A_{2n}(r,v) \exp [-\frac{1}{D}
(\phi(x)+\frac{\bar p^2}{2})]$ and $A_{2n}$ is defined
in~\eqref{eq:AnsatzPs}. Truncating this series at order $\tau^8$, we
show in Fig.~\ref{fig:pade}a that the truncation seems well behaved
for $\tau=0.01$, yielding a precise quantitative agreement with the
stationary distribution obtained numerically. Note that the
distribution develops symmetric humps which highlight the
non-Boltzmann nature of the stationary state. For $\tau=0.2$,
Fig~\ref{fig:pade}c shows that successive orders of the expansion are
typical of an asymptotic series: each order in $\tau$ contributes a
larger amount than the previous one with large positive and negative
oscillations. This need not mean that the full series fails at
capturing the steady state, but simply that finite sums yield poor
approximations of the full series. An alternative procedure to
estimate a series out of the asymptotic series of partial sums is to
use Borel resummation. To do so, we use the method of Pade
approximants. We begin by introducing the Borel transformed series
\begin{equation}
  B_N(r,\tau)=\sum_{n=0}^N \frac{\bar A_{n}(r)}{n!} \tau^{n}
\end{equation}
and we approximate $B_N$ as rational fractions
$P_N(r,\tau)=\frac{Q_N({r,\tau})}{R_N({r,\tau})}$, where $Q_N$ and
$R_N$ are polynomials in $\sqrt\tau$ of order $N$ such that $P_N$ and
$B_N$ coincide up to order $\tau^{N}$. The series
\eqref{eq:seriesmarginal} can then be evaluated from the Laplace
transform of $P_N$: $\hat P_N(r,\tau)\equiv \int_0^\infty dt P_N(r,t
\tau)e^{-t }$. The limit of $\hat P_N$ as $N\to \infty$ is called
the Borel resummation of $P_s$~\cite{serone2017power}. We first show
in Fig.~\ref{fig:pade}b that, when the series
\eqref{eq:seriesmarginal} is well approximated by its partial sum, it
is (reassuringly) also well described by its truncated Borel sum. More
interestingly, for $\tau=0.2$, when the partial sums seem to diverge,
the Borel resummation of~\eqref{eq:seriesmarginal} still agrees
quantitatively with numerical estimates of the steady-state
distribution, as shown in Fig.~\ref{fig:pade}d.


\subsection{Approximate dynamics: unified colored noise and Fox approximations}\label{subsec:Fox}

Approximate treatments of the dynamics~\eqref{eq:dyn_active_g} have
been proposed in the past, based on schemes initially developed for
non-interacting particles. They are usually referred to as the unified
colored-noise approximation (UCNA)~\cite{Hanggi:87, Cao:93} and the
Fox theory~\cite{Fox:86a, Fox:86b}. These approximations were first
motivated by the description of the fluctuations of the radiation in
the dye laser~\cite{Jung:88, Cao:93}. They have, in particular, been
used to determine approximations of the stationary distribution of
dynamics~\eqref{eq:dyn_active_g} and of mean first-passage
times~\cite{Fox:88, Bray:89, bray1990path, Jung:95}.

The UCNA consists in neglecting the left-hand side in Eq.~\eqref{eq:dyn_active_g}:
\begin{equation}\label{eq:dyn_ucna}
	\mM_{i\alpha, j\beta} \dot r_{j\beta} = - \p_{i\alpha} \Phi + (2D)^{1/2} \eta_{i\alpha} .
\end{equation}
where $\p_{i\alpha} = \p / \p r_{i\alpha}$, and we have introduced the
tensor $ \mM_{i\alpha, j\beta} = \delta_{ij} \delta_{\alpha\beta} +
\tau \p^2_{i\alpha, j\beta} \Phi $. As a result, the dynamics is now
Markovian for the particle positions, at the cost of a
position-dependent damping matrix $\mM$. The derivation
of~Eq.~\eqref{eq:dyn_ucna} yields an equation which is to be
understood with the Stratonovich convention~\cite{Jung:95}. It follows
that the associated Fokker-Planck equation for $ P ( \cur{ \br_i}, t )
$ reads
\begin{equation}\label{eq:fpe_ucna}
	\begin{aligned}
		\p_t P &= \p_{i\alpha} \pntg{ P \mM^{-1}_{i\alpha, j\beta} \p_{j\beta} \Phi }
		\\
		& \quad + D \p_{i\alpha} \brtg{ \mM^{-1}_{i\alpha, j\beta} \p_{k\gamma} \pntg{ \mM^{-1}_{k\gamma, j\beta}  P } } ,
	\end{aligned}
\end{equation}
where $\mM^{-1}$ denotes the inverse of $\mM$. A simple derivation of
the corresponding stationary distribution is detailed in
Appendix~\ref{app:AOUP}, and leads to~\cite{Maggi:15,marconi2016velocity}
\begin{equation}\label{eq:Ps_ucna}
	P_\sS(\cur{ \br_i}) \sim \exp \brt{ - \f{\Phi}{D} - \f{\tau \pnt{ \nabla_i \Phi }^2 }{2 D} } \abs{ \det \mM } .
\end{equation}
The stationary distribution differs from the equilibrium Boltzmann
distribution $ \ee^{ - \Phi / D } $, which is recovered in the
$\tau\to 0$ limit. The distribution~\eqref{eq:Ps_ucna} can be
simplified to yield~\eqref{eq:Ps_eff}, showing it to coincide with our
perturbative treatment to first order in $\tau$. Note, however, that
Appendix~\ref{app:AOUP} shows that UCNA maps the AOUPs dynamics onto
an \textit{equilibrium one}, satisfying detailed balance with respect
to the steady-state distribution~\eqref{eq:Ps_ucna}. As such, it will
be unable to capture any non-equilibrium dynamical effects, from the
emergence of currents in ratchet potentials to non-vanishing entropy
production.

The Fox theory uses projection methods to derive an approximate
Fokker-Planck description of the dynamics of
AOUPs~\cite{faetti1988projection,Fox:88}. In the spirit of this
approach, we show in Appendix~\ref{app:AOUP} that the approximate
Fokker-Planck equation of $N$ interacting AOUPs reads, within the
framework of Fox theory,
\begin{equation}\label{eq:fpe_fox}
	\p_t P = \p_{i\alpha} \pnt{ P \p_{i\alpha} \Phi } + D \p^2_{i\alpha, j\beta} \pntg{ \mM^{-1}_{i\alpha, j\beta} P } .
\end{equation}
Note that this Fokker-Planck equation differs from that
of~\cite{Brader:15}, as commented
by~\cite{rein2016applicability}. Again, both~\eqref{eq:fpe_fox} and
ref~\cite{Brader:15} approximate the dynamics of AOUPs by equilibrium
processes, albeit with different space-dependent mobilities and
diffusivities.

Comparing~\eqref{eq:fpe_fox} with the UCNA~\eqref{eq:fpe_ucna} show
both Fokker-Planck equations to share the same probability currents up
to a multiplicative factor given by $\mM^{-1}$. These two
approximations thus yield the same stationary
distribution~\eqref{eq:Ps_ucna}, which agrees quantitatively with our
perturbative result to first order in $\tau$. However, both Fox and
UCNA fail beyond this order. This may be seen directly by considering
the order $\tau^{2}$ for a particle in one dimension, for which UCNA
and Fox predict
\begin{align}
\label{eq:Ps_ucna_2}
	P_\sS(\cur{ \br_i}) \sim \exp \brt{ - \f{\Phi}{D} - \f{\tau \pnt{\Phi' }^2 }{2 D} +\tau \Phi^{''}-\frac{\tau^{2} (\Phi^{''})^2}{2}+o(\tau^{2})}
\end{align}
which differs from our systematic derivation presented in
Sec.~\ref{subsec:SM-1particle-1d}. Finally, we stress again that these
approximation schemes both map the dynamics of AOUPs onto equilibrium
ones and are thus unable to capture genuine non-equilibrium effects,
unlike---as shown in Sec.~\ref{sec:TRS}---the expansion presented in
Sec.~\ref{sec:smalltau1}.

\subsection{Spatially-varying activity}\label{sec:varying}

A marked difference between active and passive particles is that
kinetic parameters, like the particle mobility, do not impact the
steady-state distribution of passive particles, whereas they
generically matter for active particles. The prototypical example is
that of a spatially dependent propulsion speed $v(\br)$, which leads
to a non-uniform distribution $P_{\rm s}(\br)\propto \frac 1
{v(\br)}$ for RTPs~\cite{Schnitzer:93,Tailleur:08} and
ABPs~\cite{Tailleur:08}. This result can be directly generalized to
all self-propelled particles with spatially-varying self-propulsion
speeds and isotropic Markovian reorientation dynamics, whose
master-equation reads~\cite{Tailleur:13}
\begin{equation}~\label{eq:MEvofr}
  \partial_t P(\br, \boldsymbol{\theta}) =- \nabla \cdot [ v(\br) {\bf u} (\boldsymbol\theta) P(\br, \boldsymbol\theta)] + \Theta P(\br, \boldsymbol\theta)\;,
\end{equation}
where $\boldsymbol\theta$ is a $d-1$ angular vector parametrizing the
$d-1$-sphere in $d$ spatial dimensions, ${\bf u}(\boldsymbol{\theta})$ the
corresponding unit vector, and $\Theta$ is the operator accounting for
the reorientation dynamics. Any isotropic reorientation process admits
a uniform distribution over the sphere as a steady-state. Up to
normalization issue, $P(\br, \boldsymbol\theta)\propto\frac 1
{v(\br)}$ is thus a steady-state solution of~\eqref{eq:MEvofr}, which
leads to a marginal in space $P_s(\br) \propto\frac 1 {v(\br)}$. Note
that this accumulation in slow regions remains valid in the presence
of translational diffusion, but the precise form of the steady-state
now depends on the reorientation dynamics~\cite{Cates2015MIPS}.

How this result generalizes to more complex dynamics of the
self-propulsion velocity remains an open question. As we show below,
AOUPs with position-dependent $\tau(\br)$ and $D(\br)$ also
generically have non-uniform steady-states, which we characterize. We
 consider the following dynamics
\begin{equation}\label{eq:AOUPSQS}
  \begin{aligned}
    \dot \br &= \bv\\
    \tau(\br) \dot \bv &= -\bv + \sqrt{2 D(\br)} \eta
  \end{aligned}
\end{equation}
where $\tau(\br)$ and $D(\br)$ are positive functions.  The
corresponding master equation for the probability density
$P(\br,\bv;t)$ is given by
\begin{equation}\label{eq:ME-SV}
  \partial_t P(\br,\bv;t) = - \nabla \cdot (\bv P) + \nabla_{\bv} \cdot \Big(\frac \bv \tau P + \frac D {\tau^2} \nabla_\bv P\Big)\;.
\end{equation}
Interestingly, introducing $\gamma\equiv \tau^{-1}$ and
$T\equiv \frac{D}{\tau}$ maps this problem onto the dynamics of
colloidal particles with inhomogeneous temperature and damping, a
problem which has attracted a lot of attention in the
past~\cite{VK1988IBM,van1988diffusion,bringuier2007colloid,lau2007state,yang2013brownian,lim2019homogenization}. It
is then a simple exercise to check that varying $\tau(\br)$ and
$D(\br)$ while keeping $T=D/\tau$ uniform leads to a Maxwellian
steady-state $P_s(\br,\bv)\propto \exp(-\frac{\bv^2}{2 T})$ and hence
to a uniform distribution in position space $P_s(\br)$. Under more
general conditions, and somewhat surprisingly, Eq.~\eqref{eq:ME-SV}
does not seem to admit simple steady-state solutions. As we show next,
for slowly varying $\tau(\br)$ and $D(\br)$, one can nevertheless show
the steady-state distribution to be given by
$P_s(\br)\propto 1/T(\br)=\tau(\br)/D(\br)$.

\begin{figure}
  \centerline{\includegraphics{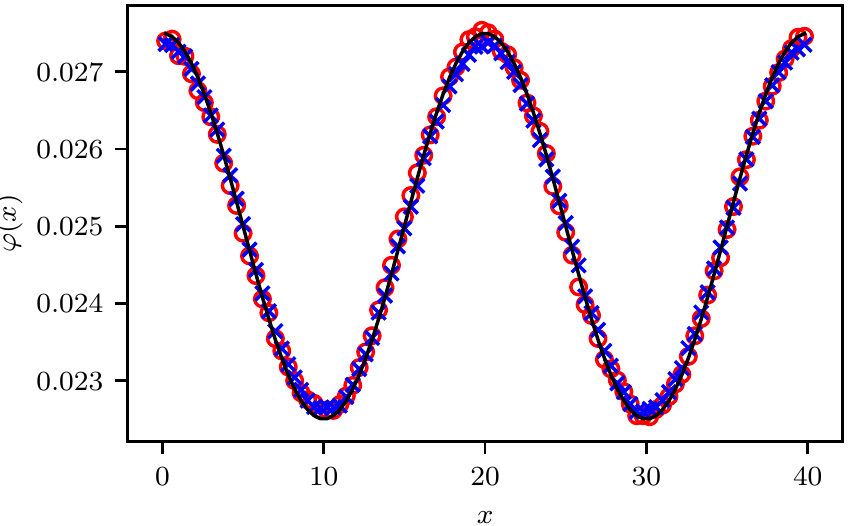}}
  \caption{Steady-state distributions of AOUPs evolving in 1d with
    $D=1$ and $\tau(x)=1+\varepsilon\cos(4 \pi x/L)$ (blue crosses)
    and with $\tau=1$ and $D=[1+\varepsilon\cos(4 \pi x/L)]^{-1}$ (red
    circles), compared with the theoretical
    prediction~\eqref{eq:SSvarying} (black line). Parameters: $L=40$,
    $\varepsilon=0.1$}\label{fig:SSvarying}
\end{figure}

Integrating~\eqref{eq:ME-SV} over ${\bf v}$ leads to the continuity
equation
\begin{equation}\label{eq:tauofrss}
  \partial_t P(\br,t) = -\nabla \cdot {\bf j}(\br,t)\;,
\end{equation}
with ${\bf j}(\br,t)=\int d{\bv} \bv
P(\br,\bv)$. Multiplying~\eqref{eq:ME-SV} by $\bv$ and integrating
over $\bv$ then leads to
\begin{equation}\label{eq:currentofr}
  \partial_t {\bf j}(\br,t) = -\nabla \cdot \bbar q(\br,t) - \frac 1 {\tau(\br)} {\bf j}(\br,t)\;,
\end{equation}
where $\bbar q=\int d\bv (\bv \otimes \bv) P(\br,\bv)$ is a
second-order tensor characterizing the local orientation
field. Finally, the dynamics of $\bbar q$ is given by
\begin{equation}\label{eq:bbarq}
  \partial_t {\bbar q_{\alpha \beta}}(\br,t) = -\nabla_\gamma \cdot \bbbar \chi_{\gamma \alpha \beta}(\br,t) - \frac {2\bbar q_{\alpha \beta}(\br,t)} {\tau(\br)} +\frac{2 D(\br) P(\br) \delta_{\alpha \beta}}{\tau(\br)^2} \,,
\end{equation}
where $\bbbar \chi=\int d\bv (\bv\otimes\bv \otimes \bv) P(\br,\bv)$
is a third-order tensor. As long as $\tau(\br)$ is bounded, the fields
$\bbar q(\br)$ and ${\bf j}(\br)$ are slaved to the density
$P(\br)$ for time-scales $t\gg {{\rm max}
  [\tau(\br)]}$. On such time-scales, assuming small gradients of
$D(\br)$ and $\tau(\br)$, Eq.~\eqref{eq:bbarq} leads to
\begin{equation}\label{eq:bbarqsol}
  {\bbar q_{\alpha\beta}}(\br,t)=\frac{D(\br)}{\tau(\br)} P(\br) \delta_{\alpha\beta}+{\cal O}(\nabla)\;.
\end{equation}
In turn, this shows the local current to take the form
\begin{equation}\label{eq:currentsol}
  {\bf j}(\br,t)=- \tau(\br) \nabla  \Big[\frac{D(\br)}{\tau(\br)} P(\br)\Big]+{\cal O}(\nabla^2)\;.
\end{equation}
Finally, this leads, to second order in gradient, to a diffusive
dynamics for $P(\br)$:
\begin{equation}\label{eq:diffappnonunif}
  \partial_t P(\br,t) = \nabla \cdot \Big[ \tau(\br) \nabla  \Big( \frac{D(\br)}{\tau(\br)} P(\br,t)\Big)\Big]
\end{equation}
Up to normalization issues, the steady-state is then given by
\begin{equation}\label{eq:SSvarying}
  P_s(\br,t) \propto \frac{\tau(\br)}{D(\br)}=\frac 1 {T({\bf r})}
\end{equation}
Note that, in this approximation, the current ${\bf j}$ vanishes in
the steady state and $\bbar q$ is uniform in space: the spatial
variations of the statistics of $\bv$ are compensated by those of
$P_s(\br)$. Eq.~\eqref{eq:SSvarying} is illustrated numerically in
Fig.~\ref{fig:SSvarying}. Eq.~\eqref{eq:diffappnonunif} also shows
that, on the time and space scales relevant to the fast variable
treatments and to the gradient expansion, AOUPs evolving with
dynamics~\eqref{eq:AOUPSQS} are equivalent to passive particles evolving
under an It\=o-Langevin dynamics
\begin{equation}\label{eq:SSVAdiffapp}
  \dot \br = D(\br) \nabla \log\tau(\br) + \sqrt{2 D(\br)}\boldsymbol{\eta}\;.
\end{equation}

Note that, without any approximation, Eq.~\eqref{eq:currentofr} shows
that, in the steady state, a spatially asymmetric periodic modulation of
$D(\br)$ and $\tau(\br)$ along one space direction cannot lead to a
non-vanishing current. Consider indeed one-dimensional modulations of
$D(\br)$ and $\tau(\br)$ along, say, the $x$
direction. Equation~\eqref{eq:tauofrss} shows the current to be
uniform in the steady state ${\bf j}=\bar{j}{\bf u}_x$. By symmetry,
Eq.~\eqref{eq:currentofr} becomes in steady-state
\begin{equation}
  -\partial_x \bbar q_{xx} = \frac{1}{\tau(x)} \bar j.
\end{equation}
Integrating over one spatial period then leads to $\bar j=0$ since
$\tau^{-1}(x)>0$ and $q_{xx}$ is periodic. This is a surprising
exception to the ratchet physics: breaking space and time symmetry may
in more general cases lead to a vanishing current. Note that this
extends to colloidal particles: an asymmetric modulation of the
temperature along a single space direction does not lead to a steady
current~\cite{VK1988IBM}. We stress that these results hold for
non-interacting particles; pairwise forces may alter this conclusion,
both for active~\cite{stenhammar2016} and passive
particles~\cite{VK1988IBM}.

\section{Motility-Induced Phase Separation}\label{sec:MIPS}

Motility-induced phase separation is a collective phenomenon observed
in self-propelled particles whose self-propulsion is hindered at high
density. The particles undergo a phase separation leading to the
emergence of dense phases in the absence of attractive forces, at odds
with the requirements for the emergence of cohesive phases in passive
systems. MIPS has been reported in experiments on self-propelled
colloids~\cite{Speck:13,van2019interrupted} and
bacteria~\cite{liu2019self}, where it led to a coexistence between
dense arrested phases and a dilute, active, disordered gas. More
recently, MIPS has been reported in a polar
liquid~\cite{geyer2019freezing} where a slow-down at high density led
to the emergence of a dense macroscopic phase, which is reminiscent of
traffic-jam
physics~\cite{chowdhury2000statistical,helbing2001traffic,nagatani2002physics}.

Self-propelled particles with persistent, non-Gaussian noises have
been shown to undergo MIPS when interacting via purely repulsive
pairwise
forces~\cite{Fily:12,Redner:13,Speck:13,stenhammar:2014:SM,wysocki:2014:EPL}.
The resulting phase separation share similarities with the one
observed for particles interacting via quorum-sensing, whose swimming
speed depends on (and decreases with) the local density of
particles~\cite{Tailleur:08,Tailleur:13,Solon:2015:EPJST}. These
sytems nevertheless display interesting differences in their
mechanical and thermodynamical
properties~\cite{Solon:15b,Solon2018:NJP}, in
particular regarding the internal structure of the liquid
phase~\cite{Tjhung:2018:PRX,caporusso2020micro,shi2020self}. A similar
phase-separation scenario has been reported for AOUPs interacting via
pairwise repulsive forces~\cite{Fodor2016PRL,maggi2020universality},
as well as for resembling kinetic Monte Carlo
dynamics~\cite{Levis:2014:PRE,Klamser:2018}. This further highlights
that, despite their simplicity, AOUPs retain the qualitative features
of self-propelled particles at the level of collective dynamics, as
was established, for instance, for the transition to collective
motion~\cite{Dossetti2015PRL}. In this section, we bring our knowledge
on MIPS in AOUPs up to par with their non-Gaussian counterparts.

We derive in section~\ref{sec:MIPSPFAPs} the collective hydrodynamics
of $N$ AOUPs interacting via pairwise repulsive forces and analyze
MIPS in this context. We highlight the similarities with the case of
non-Gaussian active particles. In particular, we show that the
dynamics of the density field is driven by the divergence of a
generalized stress tensor. Its scaling analysis can then be used to
predict a linear instability at large enough persistence, which arises
from the decrease of the flux of `active impulse'---a concept
introduced in~\cite{Fily2017JPA} and discussed in~\ref{sec:AI}---as the local
density of particles increases. Furthermore, an equation of state for
the pressure of AOUPs interacting via pairwise forces can be
established. Finally, as for ABPs and
RTPs~\cite{Speck:13,Takatori2014PRL,paliwal2018chemical,Solon2018:NJP,de2019active,bickmann2019predictive},
a first-principle theory of the phase diagram of AOUPs interacting via
pairwise forces remains an open problem. We then show in
section~\ref{sec:MIPSQSAPs} that AOUPs can also undergo MIPS due to
quorum-sensing interactions, when the latter make their persistence
times $\tau$ and noise amplitude $D$ depend on the local density of
particles. In practice, MIPS is seen when $D/\tau$ decreases
sufficiently rapidly as the local density of particles increases.

\subsection{Pairwise repulsive forces}
\label{sec:MIPSPFAPs}
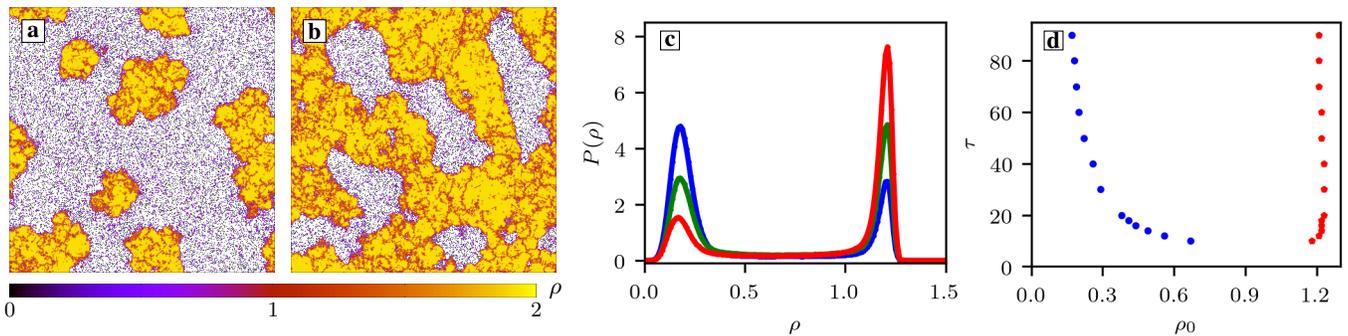
\begin{figure*}
  \begin{tikzpicture}
    \path (0.26,0) node {\includegraphics[height=100pt]{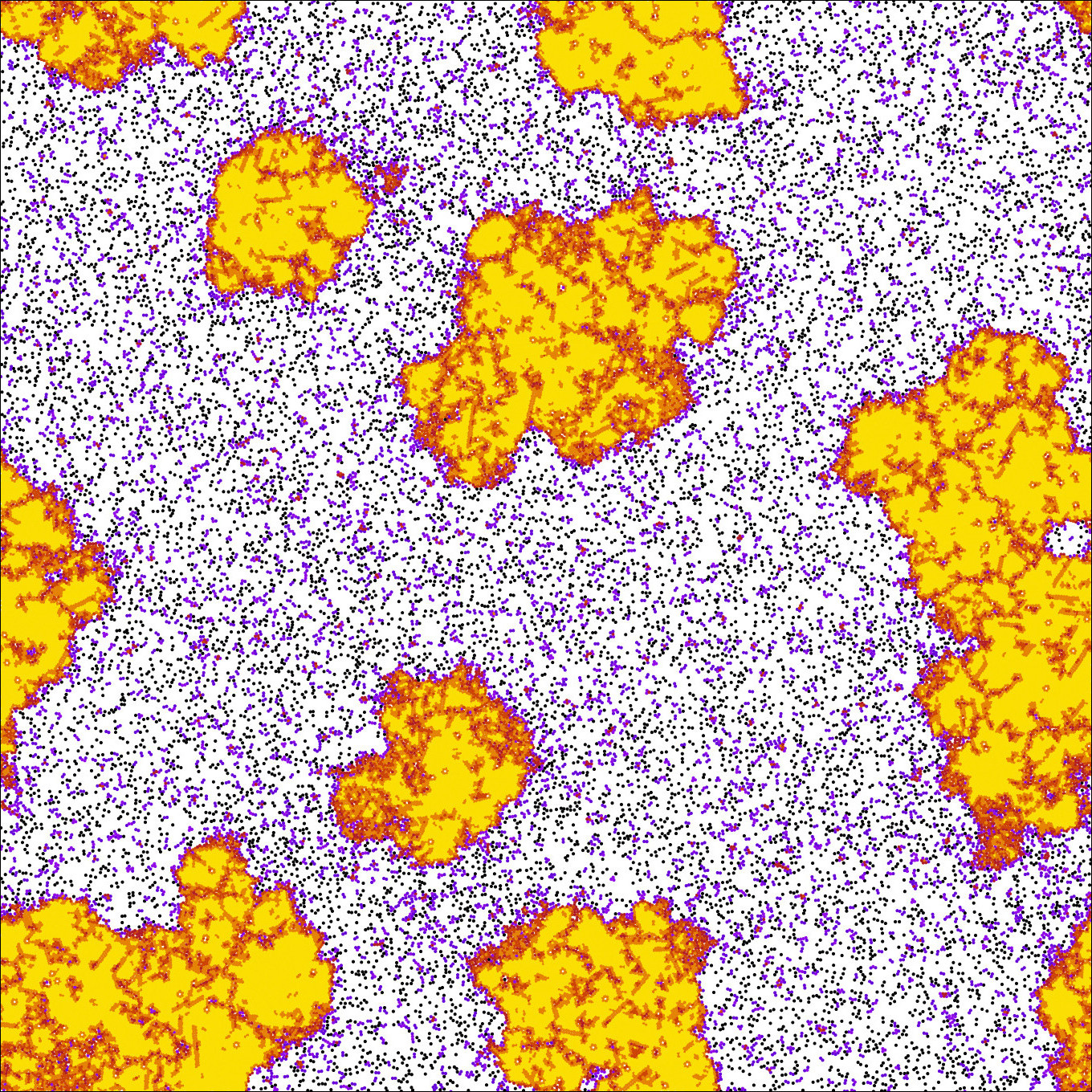}};

    \draw[fill=white] (-1.6+.26,1.6) rectangle (-1.3+.26,1.3);
    \draw (-1.45+.26,1.45) node {\bf a};

    \path (4,0) node {\includegraphics[height=100pt]{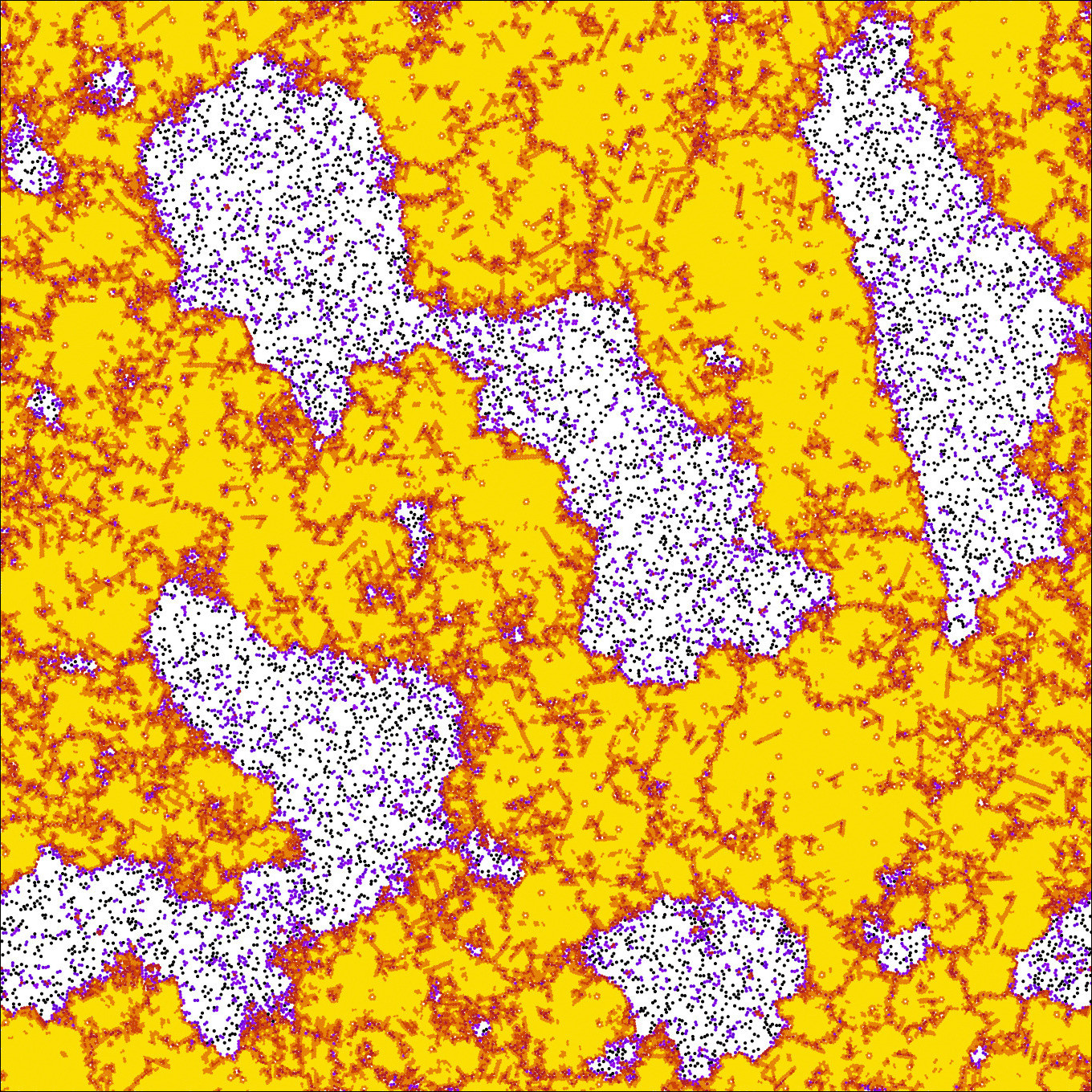}};
    \draw[fill=white] (-1.6+4,1.6) rectangle (-1.3+4,1.3);
    \draw (-1.45+4,1.45) node {\bf b};

    \path (2,-2) node {\includegraphics[angle=270,totalheight=5pt,width=7cm] {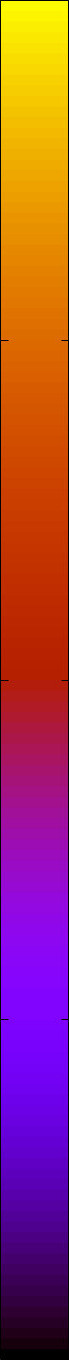}};
    \def\x{3.5}
    \def\y{-2.25}
    \draw (5.75,\y+.25) node[scale=1] {$\rho$};
    \draw (2+\x,\y) node[scale=.9] {$2$};
    \draw (2,\y) node[scale=.9] {$1$};
    \draw (2-\x,\y) node[scale=.9] {$0$};

    \def\xx{8.6}
    \path (\xx,-0.5) node {\includegraphics{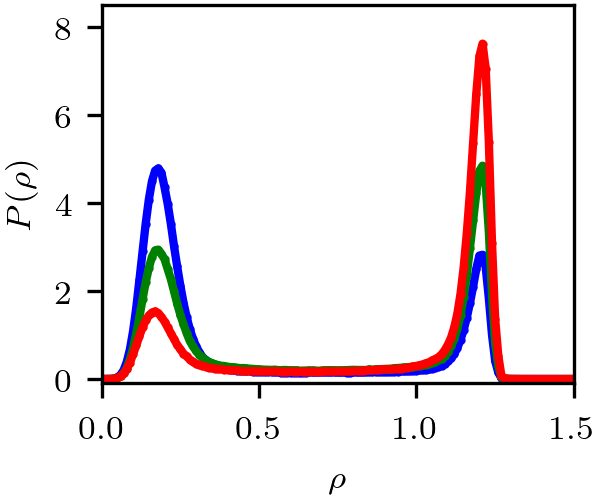}};
    \draw[fill=white] (-1.45+\xx,1.45) rectangle (-1.2+\xx,1.2);
    \draw (-1.325+\xx,1.325) node {\bf c};

    \def\xxx{13.7}
    \path (\xxx,-0.5) node {\includegraphics{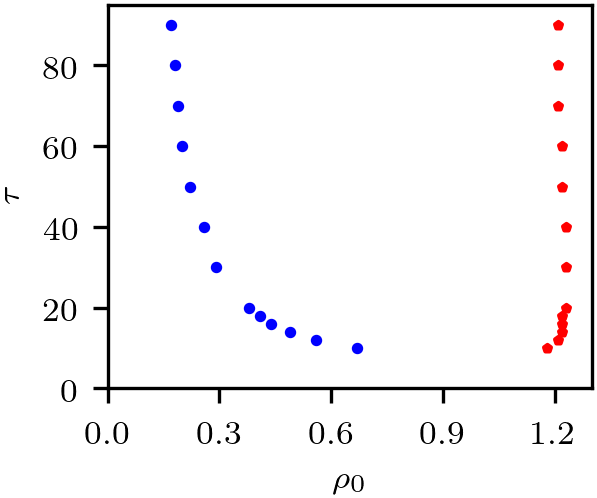}};
    \draw[fill=white] (-1.45+\xxx,1.45) rectangle (-1.2+\xxx,1.2);
    \draw (-1.325+\xxx,1.325) node {\bf d};

  \end{tikzpicture}
  \caption{Simulations of $N$ AOUPs evolving with
    dynamics~\eqref{eq:PFAPs} and~\eqref{eq:PFAPs2} and interacting
    via potential~\eqref{pot:WCA} in a 400x400 domain with periodic
    boundary conditions. Parameters: $D=10$,
    $\varepsilon=r_0=\mu=1$. In panels {\bf (a)} and {\bf (b)},
    snapshots taken after a time $t=10\,000$ show the occurence of
    motility-induced phase separation for $\tau=90$. The average
    densities are $\rho_0=0.5$ and $\rho_0=0.9$, respectively. Varying
    the overall density alters the size of the dense and dilute
    phases, but leaves their respective density unchanged. This can be
    seen from panel {\bf (c)}, which presents histograms of the local
    density measured in boxes of size 10x10. The three curves
    correspond to $\rho_0=0.50,\,0.70,\,0.90$. Finally, the phase
    diagram shown in panel {\bf (d)} is obtained by measuring the
    densities of the dilute and dense phases in
    simulations with an average density $0.9$ and different values of $\tau$. The densities are estimated from the maxima of histograms
    obtained as in {\bf (c)}.}\label{fig:MIPSPF}
\end{figure*}

We consider $N$ AOUPs interacting via purely repulsive forces
$\bff_{ij}\equiv -\nabla_i V(\br_i-\br_j)$, where $V$ is a
pairwise Weeks-Chandlers-Andersen (WCA) potential given by
\begin{equation}\label{pot:WCA}
V(r_{ij}=|\br_i-\br_j|) =  \varepsilon \left[ \left(\frac{{r_0}}{r_{ij}}\right)^{12}-\left(\frac{{r_0}}{r_{ij}}\right)^6\right]+\frac\varepsilon 4\;,
\end{equation}
when $r_{ij}<2^{1/6}{r_0}$ and $V=0$ otherwise. The dynamics of
the system read
\begin{align}
  \dot \br_i&=\bv_i + \mu \sum_j \bff_{ij}\label{eq:PFAPs}\\
  \tau \dot \bv_i&= -\bv_i + \sqrt{2 D}\bet_i \label{eq:PFAPs2}
\end{align}
For large enough $\tau$, MIPS is observed and particles self-organize
into dense arrested clusters which coexist with a dilute active gas
(See Fig.~\ref{fig:MIPSPF}).

\subsubsection{Hydrodynamic equations and generalized stress tensor}
\label{sec:AI}
To account for the underlying linear instability, we derive a
hydrodynamic equation for the density field $\rho(\br,t)$, defined as
\begin{equation}
 \rho(\br,t)=\langle \hat \rho(\br,t) \rangle\quad\mbox{with}\quad  \hat \rho(\br,t) =  \sum_{i=1}^N \delta[\br-\br_i(t)],
\end{equation}
where the average is taken with respect to the realization of the
microscopic noises $\boldsymbol{\eta}_i$. Following step-by-step the
path laid out in~\cite{Solon2018:NJP} for ABPs, and thus omitting
technical details, one finds that the dynamics of the density field is
driven by the divergence of a current, which is given in terms of the
divergence of a local `stress tensor' $\sigma$:
\begin{align}\label{eq:dynrhobar}
  \dot {\rho} &= - \nabla \cdot \boldsymbol J;\qquad \boldsymbol J = \mu \nabla \cdot \sigma\;.
\end{align}
(The mechanical interpretation of $\sigma$ is discussed in
Section~\ref{PFAOUPs:EOS}.)  In noise-free, overdamped systems like
the one considered here, the current in Eq.~\eqref{eq:dynrhobar} is
simply the particle mobility multiplied by the local force density. It
may thus come as a surprise that the latter can be written as the
divergence of a local tensor, despite the active,
momentum-non-conserving nature of the particles.

Let us first note that the stress tensor can be split between an
active and a passive part:
\begin{equation}
  \sigma\equiv \sigma^{\rm act} + \sigma^{\rm IK}\;.\\
\end{equation}
The contribution of the pairwise forces to the stress tensor is
captured by $\sigma^{\rm IK}(\br)$, which was introduced by Irving and
Kirkwood~\cite{Irving-Kirkwood-4:1950}, and whose divergence is the
local force density exerted at position $\br$:
\begin{align}\label{sigmaIK}
  \sigma^{\rm IK}_{\alpha\beta}(\br)&=\frac 1 2 \int d\br' \Big\{\frac{(\br - \br')_{\alpha} (\br - \br')_{\beta}}{|\br - \br'|} \frac{d V(|\br - \br'|)}{d|\br - \br'|}\nonumber\\&\int_0^1d\lambda\langle \hat\rho(\br + (1-\lambda)\br')\hat\rho(\br-\lambda\br')\rangle\Big\} \;.
\end{align}
This formula can be derived by symmetrizing the force density $\sum_{i\neq j}\nabla_i V({\bf r}_i-{\bf r}_j)=1/2 \sum_{i\neq j}[ \nabla_i V({\bf r}_i-{\bf r}_j) - \nabla_j V({\bf r}_i-{\bf r}_j)]$ and using that~\cite{pomeau2013surface,yvonprobleme}
\begin{equation}
  \begin{aligned}
    \delta(\br-\br_i)-\delta(\br-\br_j)=-\nabla_\br \cdot \Big[ (\br_i-\br_j)  \\
      \int_0^1 {\rm d} \lambda \delta[\br-\br_i+\lambda(\br_i-\br_j)] \Big]\;.
  \end{aligned}
\end{equation}
(This equality can be checked by series-expanding in powers of
$\lambda$ and integrating over $\lambda$, which rebuilds the Taylor
expansion of $\delta(\br-\br_i)-\delta(\br-\br_i+\br_i-\br_j)$ in
powers of $\br_i-\br_j$.)  Note that, in a system
translationnally invariant along $y$, $\sigma_{xx}^{\rm IK}(\br)$
simply measures the force density along $x$ exerted accross an
interface located at position $\br$, which endows Eq.~\eqref{sigmaIK}
with a more direct physical interpretation.

The second contribution to the stress tensor, $\sigma^{\rm act}(\br)$,
is defined from:
\begin{equation}\label{eq:bilanAI}
  \nabla \cdot
  \sigma^{\rm act}(\br) = \langle \sum_i \frac{\bv_i}\mu
  \delta(\br-\br_i) \rangle\;.
\end{equation}
Following again the path laid out in~\cite{Solon2018:NJP} then leads to:
\begin{align}
  \sigma^{\rm
  act}(\br)&\equiv \Big\langle \sum_i \frac{\bv_i \tau}\mu \big[\bv_i + \mu
    \sum_{j\neq i} \bff_{ij}\big]\delta(\br-\br_i) \Big\rangle\\
  &=  \Big\langle  \sum_i \dot \br_i \frac{\bv_i \tau}\mu  \delta(\br-\br_i) \Big\rangle\label{eq:actimp}\;.
\end{align}
Equation~\eqref{eq:actimp} provides a simple interpretation to
$\sigma^{\rm act}(\br)$: An AOUP whose self-propulsion force is
$\bv_i/\mu$ at time $t$ receives an average momentum $\tau \bv_i/\mu$
from the substrate in the future, called its `active
impulse'~\cite{Fily2017JPA}. The underlying physical picture is that
the active impulse measures the active force `stock' of the
particles. In Eq.~\eqref{eq:actimp}, $\sigma^{\rm act}(\br)$ thus
measures the flux of active impulse through the medium due to the
motion of the particles, much like for ABPs and
RTPs~\cite{Fily2017JPA}. Equation~\eqref{eq:bilanAI} then expresses
that any non-vanishing active force exerted in a volume results from
the difference between inward and outward fluxes of active impulse,
which measures what has been `spent' to maintain a non-zero steady
force.

The hydrodynamic equation~\eqref{eq:dynrhobar}, together with the
expressions~\eqref{sigmaIK} and~\eqref{eq:actimp} for $\sigma^{\rm IK}$ and
$\sigma^{\rm act}$ can be used both to predict the instability of a
homogeneous system at large enough persistence, as detailed in
Sec.~\ref{PFAOUPs:LI}, and to show the existence of an equation of
state for the pressure, as discussed in Sec.~\ref{PFAOUPs:EOS}.

\subsubsection{Scaling analysis and linear instability}\label{PFAOUPs:LI}
To proceed, it is natural to assume, following~\cite{Solon:2015:PRL}, that
the stress tensor $\sigma(\br)$ is well approximated by a local
equation of state $\sigma(\rho(\br))$. Consider a small periodic
perturbation $\delta \rho_q$, say along the $x$-direction, around a
mean density $\rho_0$. Equation~\eqref{eq:dynrhobar} shows it to
evolve as $\partial_t \delta\rho_{q}=-q^2 \mu \sigma'(\rho_0)
\delta\rho_q$ so that $\sigma'(\rho_0)<0$ signals the onset of a
linear instability and defines a spinodal region. To proceed further,
one could approximate the various components of $\sigma$ using local
equations of states $\sigma^{\rm IK}(\rho)$, $\sigma^{\rm act}(\rho)$,
as was done for ABPs~\cite{Solon:2015:PRL}. However, direct inspection of
their physical origin already captures most of the physics.

Much like for purely repulsive passive particles, $\sigma^{\rm
  IK}(\rho_0)$ vanishes in the low-density, non-interacting limit and
increases monotonously until it diverges at
close-packing~\cite{fisher1965bounds}. Because particles stop when the
interparticle force balances the propulsive one, the scale of
$\bff_{ij}$ is set by $v_0/\mu$, where $v_0=\sqrt{D/\tau}$ is the
typical scale of the self-propulsion speed. The integral
in~\eqref{sigmaIK} then selects the typical interaction length ${r_0}$
and the overall scaling of $\sigma^{\rm IK}$ is
\begin{equation}\label{eq:sigmaikscaling}
  \sigma^{\rm IK}(\rho_0) \sim \rho_0 \sqrt{\frac{D}\tau} \frac{{r_0}}{\mu} S
\end{equation}
where the dependence on $\rho_0$ has been formally written, for
dimensional reasons, as $\rho_0 S$, where $S$ is a dimensionless
scaling function of the rescaled density~\cite{Solon:2015:PRL}. (Below
a crowding density $\rho_c$, $S$ is well approximated by a linearly
decaying function
$S=1-\rho/\rho_c$~\cite{Fily:12,Bialke2013EPL,Solon:15b,Solon2018:NJP}). Note
that~\eqref{eq:sigmaikscaling} can be derived using~\eqref{sigmaIK}
or, equivalently, using the Virial-like formula
$\sigma^{\rm IK}\sim \langle \sum_{i\neq j} (\br_i-\br_j)\otimes
\nabla_i
V(\br_i-\br_j)\rangle$~\cite{Manning:14,epstein2019statistical}.

Contrary to $\sigma^{\rm IK}$, $\sigma^{\rm act}(\rho_0)$ is expected
to be a non-monotonous function of $\rho_0$. It first increases
linearly with the density until interactions kick in. Because active
particles collide more often with particles lying ahead than
behind~\cite{Bialke2013EPL}, the main effect of these interactions is
to \textit{lower} the flux of active impulse as the density increases,
until it vanishes at close packing, leading to a non-monotonous
variation of $\sigma^{\rm act}$. Inspection of Eq.~\eqref{eq:actimp}
shows the typical scale of $\sigma^{\rm act}$ to be
\begin{equation}\label{eq:sigmaactscaling}
  \sigma^{\rm act} \sim \rho_0 \frac{v_0^2 \tau}\mu \tilde S(\rho_0)=
  \rho_0 \frac D \mu \tilde S
\end{equation}
where $\tilde S(\rho)$ is another dimensionless scaling function.

An overall decreasing stress tensor can then be observed when the
decrease of $\sigma^{\rm act}(\rho_0)$ is strong enough to compensate
the increase of $\sigma^{\rm IK}(\rho_0)$. From
Eqs~\eqref{eq:sigmaikscaling} and~\eqref{eq:sigmaactscaling}, this is
realized when the rescaled persistence length is large enough:
\begin{equation}
  \ell_p=\frac{\sqrt{D \tau}}{{r_0}} \gg \frac{\tilde S}{S}\;.
\end{equation}
Let us highlight, once again, the similarity of this criterion with
that obtained for ABPs or RTPs~\cite{Solon:2015:PRL}.

\subsubsection{An equation of state for pressure}\label{PFAOUPs:EOS}
Finally, before turning to quorum-sensing interactions, let us note
that, as for other active particles interacting via pairwise
forces~\cite{Solon2018:NJP}, the analogy between $\sigma$ and a stress
tensor goes further than Eq.~\eqref{eq:dynrhobar}: it has a direct
mechanical interpretation. Confining AOUPs by an external potential
$V_{\rm w}$ indeed leads to
\begin{equation}\label{eq:fluxextwall}
  \boldsymbol{J}(\br)=\mu \nabla \cdot [\sigma^{\rm act}(\br)+\sigma^{\rm IK}(\br)] - \mu \rho \nabla V_{\rm w}
\end{equation}
with $\sigma^{\rm IK}$ and $\sigma^{\rm act}$ given by~\eqref{sigmaIK}
and~\eqref{eq:actimp}.  In a flux-free steady state,
Eq.~\eqref{eq:fluxextwall} shows the force density exerted by the
particles on the confining boundary, $\rho\nabla V_{\rm w}$, to be
given by the divergence of the stress tensor $\sigma$. Consider a
confining boundary parallel to $\hat e_y$.
Integrating~\eqref{eq:fluxextwall} from a point $\br_{\rm bulk}$, deep
in the bulk of the system, to infinity along $\hat e_x$ shows the
mechanical pressure exerted by the particles on the wall,
$P=\int_{x_{\rm bulk}}^{\infty} \rho(x,y) \partial_x V_w dx$, to be
given by $P=-\sigma_{xx}(\br_{\rm bulk})$. In a bulk phase-separated
system, $x_{\rm bulk}$ can be equally chosen in any of the phases,
showing that the pressure is identical in either of the coexisting
phases.

\subsection{Quorum-sensing interactions}
\label{sec:MIPSQSAPs}
From bacteria to self-propelled Janus colloids, many active particles
experience propelling forces whose statistics depend on the
composition of their environment and may thus be altered by the
presence of other nearby active particles. Such mediated interactions,
whether chemical, hydrodynamic or metabolic, can be
modelled---somewhat crudely---by quorum-sensing (QS)
interactions. These can be described using a self-propulsion whose
characteristics depend on the density of neighbouring
particles. Consider the dynamics of ABP or RTP, $\dot\br = v
\bu (\theta)$, where $\theta$ undergoes either rotational diffusion or
a Poisson jump process, respectively. Quorum-sensing interactions can
be modelled by considering a self-propulsion speed that is both a
function of the position of the particle $\br$ and a functional of the
density field $\hat \rho$: $v=v(\br,[\hat \rho])$~\cite{Solon:2015:EPJST,Solon2018:NJP,bauerle2018self}. In this context, a
homogeneous phase at density $\rho_0$ has been shown
to be linearly unstable to MIPS
whenever~\cite{Tailleur:08,Tailleur:13,Cates2015MIPS}
\begin{equation}\label{eq:oldMIPS}
  \frac{\rm d}{{\rm d}\rho_0} \log[v(\rho_0)]<-\frac 1 {\rho_0}\;,
\end{equation}
where $v(\rho_0)$ is the self-propulsion speed $v({\bf r},[\hat\rho])$ of a
particle in a homogeneous system at density $\rho_0$. The role of QS
for AOUPs with varying propulsion speed has not been studied so
far. To fill this gap, we consider $N$ AOUPs, whose dynamics are given
by $\dot \br_i = \bv_i$ and
\begin{equation}\label{eq:AOUPsQSAPs}
  \tau(\br_i,[\hat\rho]) \dot \bv_i = - \bv_i + \sqrt{2 D(\br_i,[\hat\rho])} \boldsymbol{\eta}_i
\end{equation}
where $\tau$ and $D$ are functions of $\br$ and functionals of the
density field $\hat\rho(\br)$. In practice,
we choose $\tau(\br,[\hat \rho])=\tau(\tilde\rho(\br))$ and
$D(\br,[\hat \rho])=D(\tilde\rho(\br))$ where $\tilde\rho(\br)$ is a
local smooth density field felt by a particle at position $\br$.
It is defined through
\begin{equation}
  \tilde \rho(\br) = \int d \br' K(\br-\br') \hat \rho(\br')\,,
\end{equation}
with $K(\br)$ a positive, symmetric kernel, normalized by $\int {\rm
  d}\br K(\br)=1$ and with a typical range $\ell_0$.

\subsubsection{A diffusive fluctuating hydrodynamics}
We consider a long-time,
large system-size limit in which the density field evolves over
time-scales $t\propto L^z$, where $z$ is the corresponding dynamical
exponent, so that there is a clear separation between the persistence
time $\tau$ and the time-scale of the evolution of the density
field. We thus assume that the dynamics of particle $i$ is well
described by its diffusive
approximation. Using~\eqref{eq:SSVAdiffapp}, this is given by:
\begin{equation}\label{eq:Langevinapprox}
  \dot \br_i = D(\tilde \rho(\br_i)) \nabla \log[\tau(\tilde\rho(\br_i))] + \sqrt{2 D(\tilde\rho(\br_i))}\boldsymbol{\eta_i}\;.
\end{equation}

Eq.~\eqref{eq:Langevinapprox} amounts to a diffusive approximation of
the dynamics of the $N$ interacting AOUPs by coupled Langevin
dynamics. Standard methods of stochastic
calculus~\cite{Dean:96,Tailleur:13,Solon:2015:EPJST} then allow one to derive
a stochastic evolution equation for $\hat \rho$. Altogether, the
fluctuating hydrodynamics of $N$ interacting AOUPs is given by
\begin{equation}\label{eq:QSAOUPsFH}
  \partial_t {\hat \rho}(\br,t) = \nabla \cdot \Big[ D \nabla \hat \rho + \hat \rho\tau \nabla\Big(\frac{D}{\tau}\Big)+ \sqrt{2 D \hat \rho} \Lambda(\br,t)\Big]\,,
\end{equation}
where $\Lambda(\br,t)$ is a Gaussian white-noise field of zero mean
and unit variance.

\subsubsection{Linear stability analysis of the mean-field hydrodynamics}
Equation~\eqref{eq:QSAOUPsFH} is the main result of this section: it
provides the fluctuating hydrodynamics of $N$ AOUPs interacting via QS
through $D(\tilde\rho)$ and $\tau(\tilde \rho)$. The structure of this
equation is very similar to that of RTPs and ABPs interacting via
density-dependent self-propulsion speed~\cite{Tailleur:13}, but the
relationship between the macroscopic transport parameters and the
microscopic parameters of the self-propulsion are clearly
different. The methods developed to study the collective behaviours of
RTPs and ABPs interacting via QS~\cite{Solon:2018:PRE,Solon2018:NJP}
can be directly generalized to AOUPs. First, Eq.~\eqref{eq:QSAOUPsFH}
immediately leads to an exact evolution equation for the mean density
field $\rho(\br,t)$ through:
\begin{equation}\label{eq:QSAOUPsH}
  \partial_t {\rho}(\br,t) = \nabla \cdot \Big[ \Big\langle D(\br,[\hat \rho]) \nabla \hat \rho + \hat \rho\tau(\br,[\hat \rho]) \nabla\Big(\frac{D(\br,[\hat \rho])}{\tau(\br,[\hat \rho])}\Big)\Big\rangle \Big]\,,
\end{equation}
Note that this equations is neither closed, nor local. A further
simplifying step is to implement a mean-field approximation and replace
$\langle f(\br ,[\hat \rho])\rangle $ by $f(\br,[\rho])$, yielding
\begin{equation}\label{eq:QSAOUPHMF}
  \partial_t {\rho}(\br,t) = \nabla \cdot \Big[ D(\br,[\rho]) \nabla \rho + \rho\tau(\br,[\rho]) \nabla\Big(\frac{D(\br,[\rho])}{\tau(\br,[\rho])}\Big) \Big]\,.
\end{equation}
At this stage, Eq.~\eqref{eq:QSAOUPHMF} already allows one to predict
the onset of a linear instability leading to MIPS. This is done by
computing the linear dynamics of a perturbation $\delta
\rho(\br,t)\equiv \rho(\br,t)-\rho_0$ around a homogeneous profile
$\rho(\br,t)=\rho_0$, leading to
\begin{equation}
  \partial_t \delta {\rho}(\br,t) =  D(\rho_0) \Delta \delta \rho + \rho_0 \tau(\rho_0) \left(\frac{D(\rho_0)}{\tau(\rho_0)}\right)' \Delta \delta \rho\;,
\end{equation}
where $\Delta= \nabla^2$, the prime refers to a derivative with
respect to $\rho_0$, and we defined $D(\rho_0)\equiv D(\br,[\rho_0])$
and similarly for $\tau(\rho_0)$. A Fourier transform then shows that
mode $q$ relaxes with rate $\Lambda_q$, where
\begin{equation}
  \Lambda_q=D(\rho_0) q^2 \left (1+\rho_0 \log\left(\frac{D(\rho_0)}{\tau(\rho_0)}\right)' \right)
\end{equation}
The mode is unstable whenever $\Lambda_q<0$, whence the instability
criterion
\begin{equation}\label{eq:QSAOUPspinodal}
  \left(\log \frac{D(\rho_0)}{\tau(\rho_0)}\right)' < -\frac{1}{\rho_0}\;,
\end{equation}
Equation~\eqref{eq:QSAOUPspinodal} defines the
spinodal region of AOUPs interacting via quorum-sensing; it is the
direct counterpart for AOUPs of the standard MIPS
criterion~\eqref{eq:oldMIPS} derived for ABPs and RTPs.

\begin{figure}
  \includegraphics{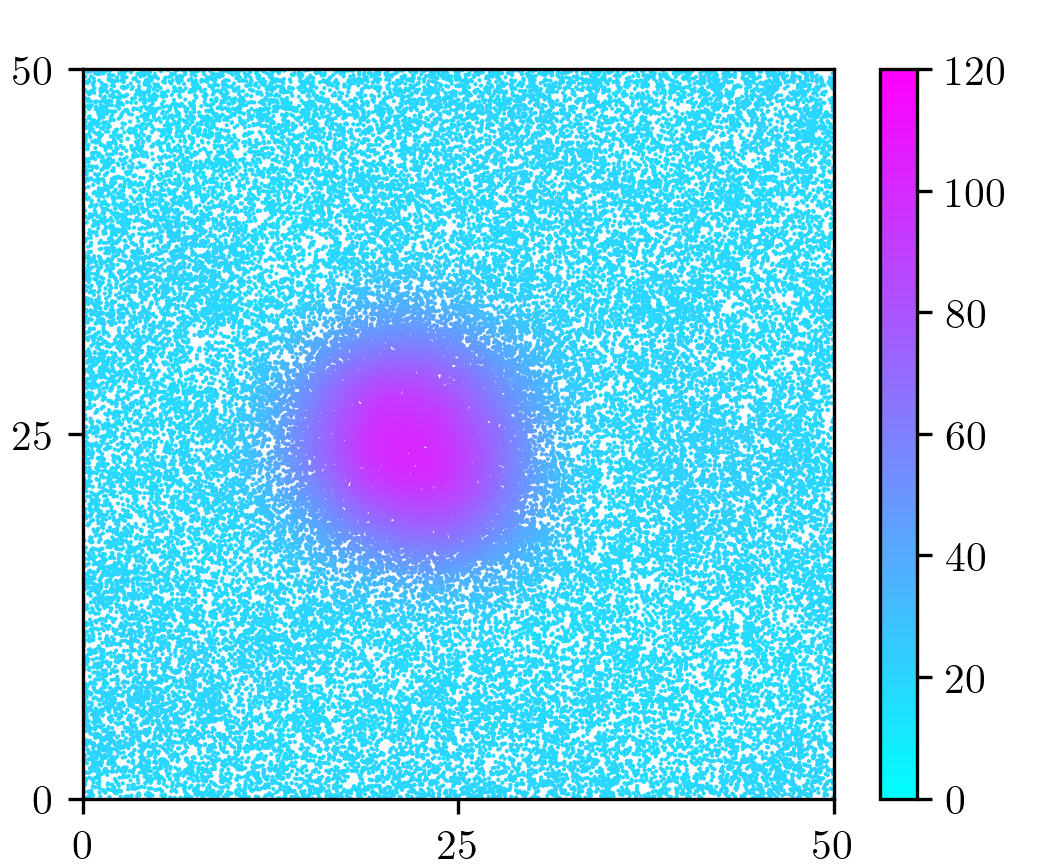}
  \caption{Snapshots of $N$ AOUPs interacting via quorum-sensing
    through Eq.~\eqref{eq:AOUPsQSAPs} with $\tau=1$ and
    $D(\br,[\hat\rho])=D_0 e^{-\lambda \phi \arctan(\tilde
      \rho(\br)/\phi)}$. The field $\tilde \rho(\br)$ measures the
    density field $\hat\rho$ averaged over a disk of radius 1 centered in
    $\br$. A linear instability will lead to MIPS whenever $\lambda
    \phi>2$, according to Eq.~\eqref{eq:QSAOUPspinodal}. Simulation
    parameters: $D_0=1$, $\rho_0=N/L^2=25$, $L=50$, $\lambda=0.05$,
    $\phi=44.72$, so that $\lambda \phi=2.23$. Color encodes the
    density averaged over a disk of radius 5. Starting from a random
    initial condition, the snapshot is taken after a time
    $t=49\,800$.}
  \label{fig:QSAOUPs}
\end{figure}

Note that Eq.~\eqref{eq:QSAOUPspinodal} predicts a density-dependent
persistence time to lead to MIPS whenever $\rho_0
\tau'(\rho_0)>\tau(\rho_0)$ whereas density-dependent tumbling rates
or rotational diffusion, which control the persistence times of ABPs
and RTPs, do not lead to any interesting collective behaviours. The
MIPS experienced by AOUPs interacting via such a density-dependent
persistence time is illustrated in Figure~\ref{fig:QSAOUPs}.

\subsubsection{Phase diagram}
The full phase diagram of ABPs and RTPs interacting via quorum-sensing
interactions has recently been predicted analytically by carrying out
one more approximation step~\cite{Solon:2018:PRE,Solon2018:NJP}. We
present here a direct application of this method to AOUPs. We first
recast~\eqref{eq:QSAOUPHMF} into
\begin{eqnarray}\label{eq:potchim}
  \partial_t {\rho}(\br,t) &=& \nabla \cdot \big[ \rho D(\br,[\rho]) \nabla g(\br,[\rho])\big]\\
    g(\br,[\rho])&=& \log\rho+\log\frac{D(\br,[\rho])}{\tau(\br,[\rho])}
\end{eqnarray}
Equation~\eqref{eq:potchim} shows that $g(\br,[\rho])$ acts as a
chemical potential. Using a second-order gradient expansion, the
non-local sampling of the density field through the kernel $K({\bf u})$ can
be written as:
\begin{equation}
  \langle \tilde \rho(\br) \rangle \simeq \rho(\br) + \frac 1 2
  \ell_0^2 \Delta \rho(\br)
\end{equation}
where $\ell_0^2=\int {\bf u}^2 K({\bf u}) {\rm d}{\bf u}$. This allows
us to expand $g(\br,[\rho])$ into
\begin{equation}
  g(\br,[\rho])=g_0(\rho(\br)) - \kappa(\rho(\br))\Delta \rho(\br)
\end{equation}
where
\begin{eqnarray}
  g_0(\rho)&=&\log\rho+\log\frac{D(\rho)}{\tau(\rho)}\;\\
  \kappa(\rho)&=&\frac{\ell_0^2}{2}\left(\frac{\tau'(\rho)}{\tau(\rho)}-\frac{D'(\rho)}{D(\rho)}\right)
\end{eqnarray}
Following the method laid out in~\cite{Solon:2018:PRE,Solon2018:NJP},
one introduces the change of variable $R(\rho)$, solution of
$R'(\rho)=\frac{1}{\kappa(\rho)}$. This allows us to write
\begin{equation}
  g(\br)=-\frac{\delta  \mathcal{F}[R]}{ \delta R(\br)}
\end{equation}
where ${\cal F}[R]=\int {\rm d}\br \big[\phi(R)+\frac{\kappa}{2R'}(\nabla
R)^2\big]$ is the generalized free energy whose local density is such that
\begin{equation}
  \frac{{\rm d} \phi(R)}{{\rm d}R}=g_0(\rho(R))
\end{equation}
The phase diagram can then be predicted, at this diffusive, mean-field
level, by carrying out a common-tangent construction on
$\phi(R)$. This was shown in \cite{Solon:2018:PRE,Solon2018:NJP} to
give quantitative agreement with microscopic simulations as soon as
$\ell_0$ is large enough that each particle interacts with many of its
neighbours.


\section{Emergence of time-reversal symmetry breaking}
\label{sec:TRS}

In this Section, we discuss observables which can be used to measure
the deviation from equilibrium of AOUPs. First, we consider the
particle current which arises spontaneously when introducing an
external asymmetric potential. We obtain its expression to leading
order based on the small-persistence-time expansion detailed in
Sec.~\ref{sec:fpe}. Then, we derive the entropy production rate
quantifying the breakdown of time-reversal
symmetry~\cite{Seifert:12}. This observable has been used
extensively as an unambigous nonequilibrium signature in systems
driven by external
fields~\cite{evans1993probability,gallavotti1995dynamical,kurchan1998fluctuation,Lebowitz,maes1999fluctuation} and
it has attracted a lot of attention recently in the field of active
matter~\cite{Speck2016,Fodor2016PRL,Marconi2017,Pietzonka2017,mandal2017entropy,Shankar2018,Chaki2018,Dabelow2019,caprini2018comment,caprini2019entropy,chaki2019effects,Grandpre2020,PhysRevE.102.022607}. Finally,
we end by discussing the symmetry of time correlations and its
relation to entropy production.

\subsection{Current and Ratchet}
\label{sec:current}
Nonequilibrium systems can sustain currents in the steady state. One
of the simplest settings for this to happen is a stochastic ratchet: a
fluctuating nonequilibrium dynamics in a spatially asymmetric
potential landscape generically leads to non-vanishing currents in the
steady state~\cite{ajdari1992mouvement,magnasco1993forced}. Recently,
several works considered active particles in asymmetric landscapes,
both
experimentally~\cite{galajda2007wall,angelani2009self,di2010bacterial,sokolov2010swimming,kumar2016sorting}
and
theoretically~\cite{wan2008rectification,Cates:09,angelani2011active,reichhardt2017ratchet,stenhammar2016,Pietzonka2019}.

\begin{figure}
\includegraphics[width=\columnwidth]{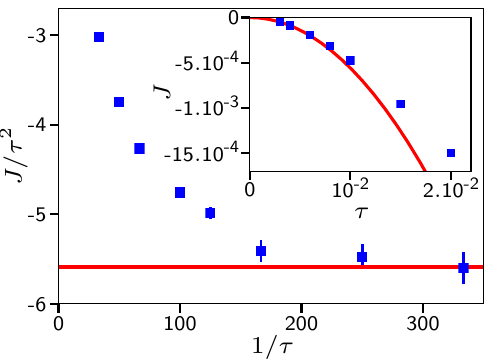}
\caption{\label{fig:J_profile}(Color online) Plot of the normalized
  current $J/\tau^2$ induced by a ratchet potential $\Phi(r)=\sin(\pi
  r/2)+\sin(\pi r)$ as a function of the inverse of the persistence
  time $\tau^{-1}$. The blue dots correspond to numerical simulations
  with error bars given by the standard deviation; the red line is our
  analytical prediction in the small $\tau$ limit obtained from
  Eq. (\ref{eq:CurrentSmooth}). In the inset, we plot $J$ as a
  function of $\tau$.  }
\end{figure}

We consider here a single AOUP on a ring of length $L$ in an
asymmetric potential of period $L$. Our interest goes to the induced
current $J=\langle p\rangle$ in the steady state. An interesting
spin-off of our computation of the stationary measure detailed in
Sec. \ref{subsec:SM-1particle-1d} and Appendix~\ref{app:SM} is a
perturbative expression for $J$. To leading order in $\tau$, $J$
reads:
\begin{equation}
\label{eq:CurrentSmooth}
J=\langle p\rangle=\tau^{2} \frac{L\int_{0}^{L}\Phi'(r)^2 \Phi^{(3)}(r)dr}
{2\int_{0}^{L}e^{\frac{\Phi}{D}}dr\int_{0}^{L}e^{-\frac{\Phi}{D}}dr}\,
+o(\tau^2)\,.
\end{equation}
We compare in Fig. \ref{fig:J_profile} the above prediction with the
results of numerical simulations of an AOUP experiencing a potential
$\Phi(r)=\sin(\pi r/2)+\sin(\pi r)$. The agreement between our
numerical simulations and Eq.~\eqref{eq:CurrentSmooth} confirms the
validity of our approximation at small $\tau$. In particular, this is
an effect that could not be captured by the UCNA or Fox approximations
described in Sec.~\ref{subsec:Fox}: both would predict $J=0$.

\subsection{Entropy production rate}
\label{sec:sig}
To probe quantitatively the nonequilibrium properties of the dynamics,
we now consider the entropy production rate $\cS$~\cite{Lebowitz,
  Seifert:12}. It is defined as the rate of (Kullback-Leibler)
divergence between the probability weights associated with a given
realization of the dynamics, $\{\br_i(t),\bp_i(t)\}$, and its time-reversed counterpart, $\{\br^\R_i(t),\bp^\R_i(t)\}$,
respectively denoted by $\cP$ and $\cP^\R$, as
\begin{equation}
	\cS = \underset{t_f\to\infty}{\lim} \f{1}{t_f} \left\langle \ln \f{\cP[\{\br_i(t),\bp_i(t)\}]}{ \cP^\R[\{\br_i^\R(t),\bp_i^\R(t)\}] } \right\rangle\;,
\end{equation}
where $t_f$ is the length of the trajectory. The entropy production
rate $\cS$ quantifies the irreversibility of the dynamics. Using
standard path-integral formalism~\cite{Onsager:53}, the trajectory
weight can be written as $ \cP \sim \ee^{ - \cA } $ , where the
dynamic action $\cA$ reads
\begin{equation}\label{eq:defaction}
  \cA = \f{1}{4D} \int_0^{t_f} \brt{ \tau \dot \bp_i + \bp_i + ( 1 + \tau \bp_j \cdot \nabla_j ) \nabla_i \Phi }^2 \dd t.
\end{equation}
We define the reversed trajectories $ \cur{ \br^\R, \bp^\R }$ in terms of the forward ones as
\begin{equation}\label{eq:rev_traj}
	\br_i^\R (t) = \br_i (t_f-t) ,
	\quad
	 \bp_i^\R (t) = - \bp_i(t_f-t) .
\end{equation}
Note that this amounts to comparing a forward trajectory with a
backward one, realized by a particle whose initial velocity is the
opposite of the final velocity of the forward trajectory, a choice
which has been debated~\cite{mandal2017entropy,caprini2018comment}.

From this definition, we deduce the action difference $ \delta \cA =
\cA - \cA^\R $ as
\begin{equation}
\label{eq:sigg}
	\begin{aligned}
		\delta \cA &= \f{1}{D} \int_0^{t_f} \cur{ \bp_i \cdot \nabla_i \Phi + \tau \brt{ \dot \bp_i \cdot \bp_i +   \nabla_i \Phi \cdot ( \bp_j \cdot \nabla_j ) \nabla_i \Phi } \right.
		\\
		& \left. \quad + \tau^2 \dot \bp_i \cdot ( \bp_j \cdot \nabla_j ) \nabla_i \Phi } \dd t ,
	\end{aligned}
\end{equation}
where $\cA^\R=\cA(\br_i^\R(t),\bp_i^\R(t))$ and we use the Stratonovich
convention. Using $ \bp_i = \dot \br_i $, it appears that the first
line in~\eqref{eq:sigg} integrates into a finite contribution $[\Phi +
  \tau \bp^2/2+\tau (\nabla \Phi)^2/2]_0^{t_f}$, which does not
contribute to $\cS$ as $t_f\to \infty$. We assume that the time and
ensemble averages coincide under the ergodicity condition, leading us
to express the entropy production rate as
\begin{equation}
	{\cal S} = - \frac{\tau^2}{D} \big\langle (\dot{\bf p}_i\cdot\nabla_i)({\bf p}_j\cdot\nabla_j)\Phi \big\rangle .
\end{equation}
Using the chain rule
\begin{equation}
	\frac{\rm d}{{\rm d}t} \big\langle ({\bf p}_i\cdot\nabla_i)^2\Phi \big\rangle = 2\big\langle (\dot{\bf p}_i\cdot\nabla_i)({\bf p}_j\cdot\nabla_j)\Phi \big\rangle + \big\langle ({\bf p}_i\cdot\nabla_i)^3\Phi \big\rangle ,
\end{equation}
and given that ${\rm d}\langle A\rangle/{\rm d}t$ vanishes in steady state for any observable $A$, we then deduce
\begin{equation}\label{eq:EP}
	\cS = \f{ \tau^2 }{2D} \avg{ ( \bp_i \cdot \nabla_i )^3 \Phi }\;.
\end{equation}
The entropy production rate vanishes when $\Phi$ is quadratic, a case
which has attracted interest in the past~\cite{Szamel:14}. In such a
harmonic trap, AOUPs have a Gaussian, Boltzmann-like distribution,
albeit with a potential-dependent
``temperature''. Equation~\eqref{eq:EP} shows that this quantitative
difference with thermal equilibrium does not imply a breakdown of
time-reversal symmetry, in the sense that detailed balance holds. The
anharmonicity of the potential can thus be used as a handle to drive
AOUPs out of equilibrium. Note that we use ${\cal S}$ solely to detect
a breakdown of time-reversal symmetry. It can also be granted a more
traditional thermodynamical meaning, as recently discussed
in~\cite{Dabelow2019,Pietzonka2019,EkehPRE2020}.

Equation~\eqref{eq:EP} is a global measure of the entropy production
 rate over the whole system. We now turn to a more detailed study of how
this entropy is locally produced and, more precisely, of the spatial
structures which are most sensitive to time-reversal symmetry
breaking.

\begin{figure*}
\includegraphics[width=\columnwidth]{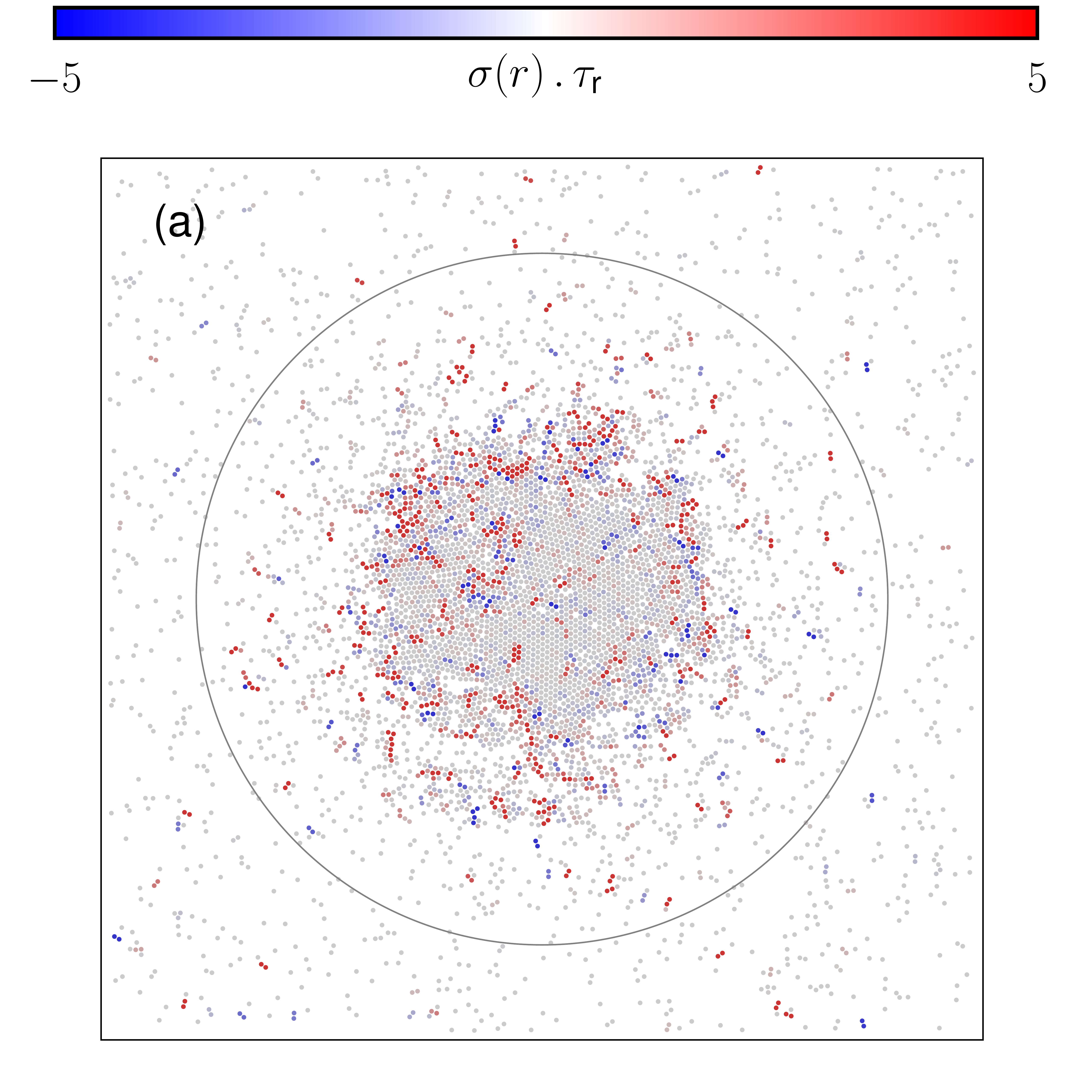}
\hfill
\includegraphics[width=\columnwidth]{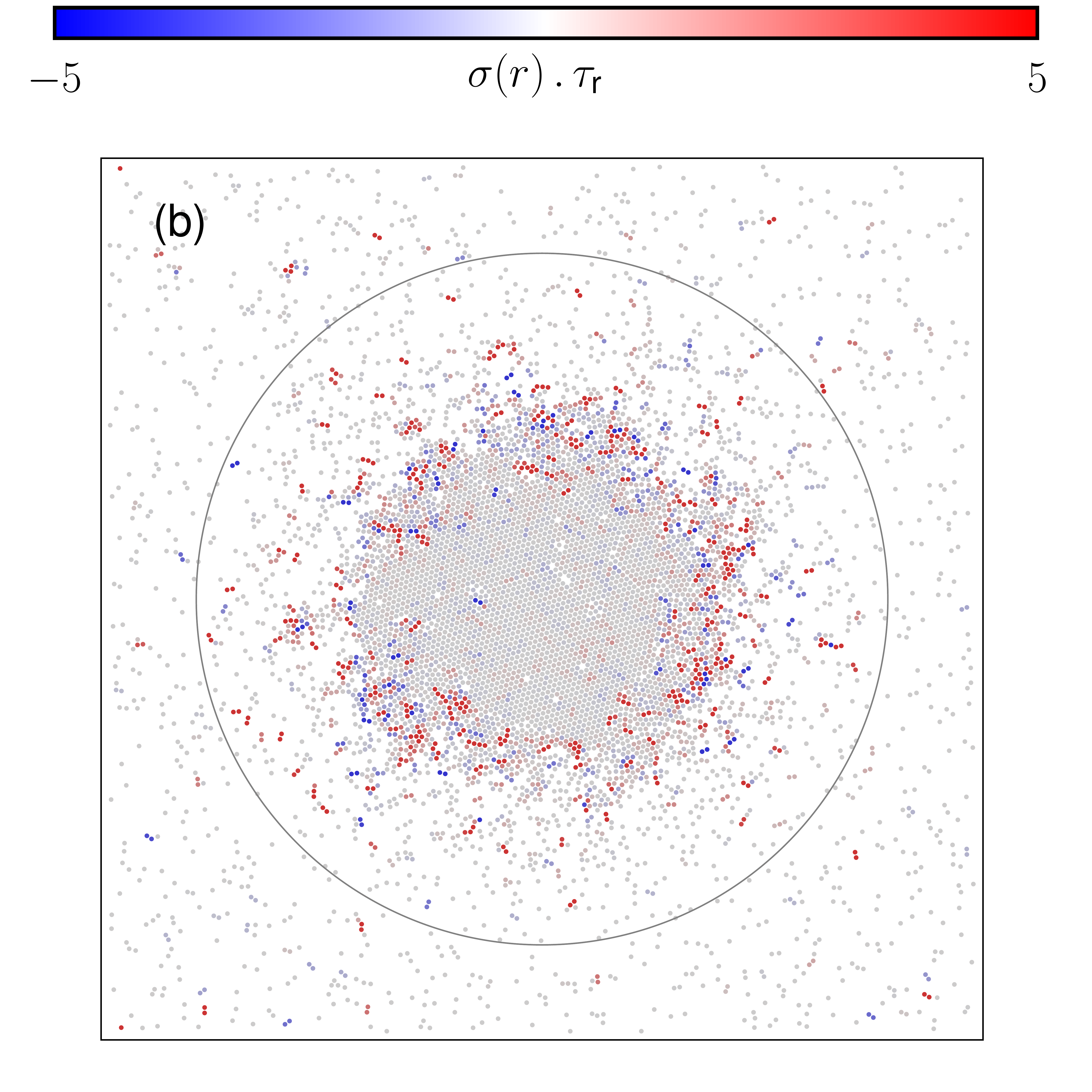}
\vskip.5cm
\includegraphics[width=\columnwidth]{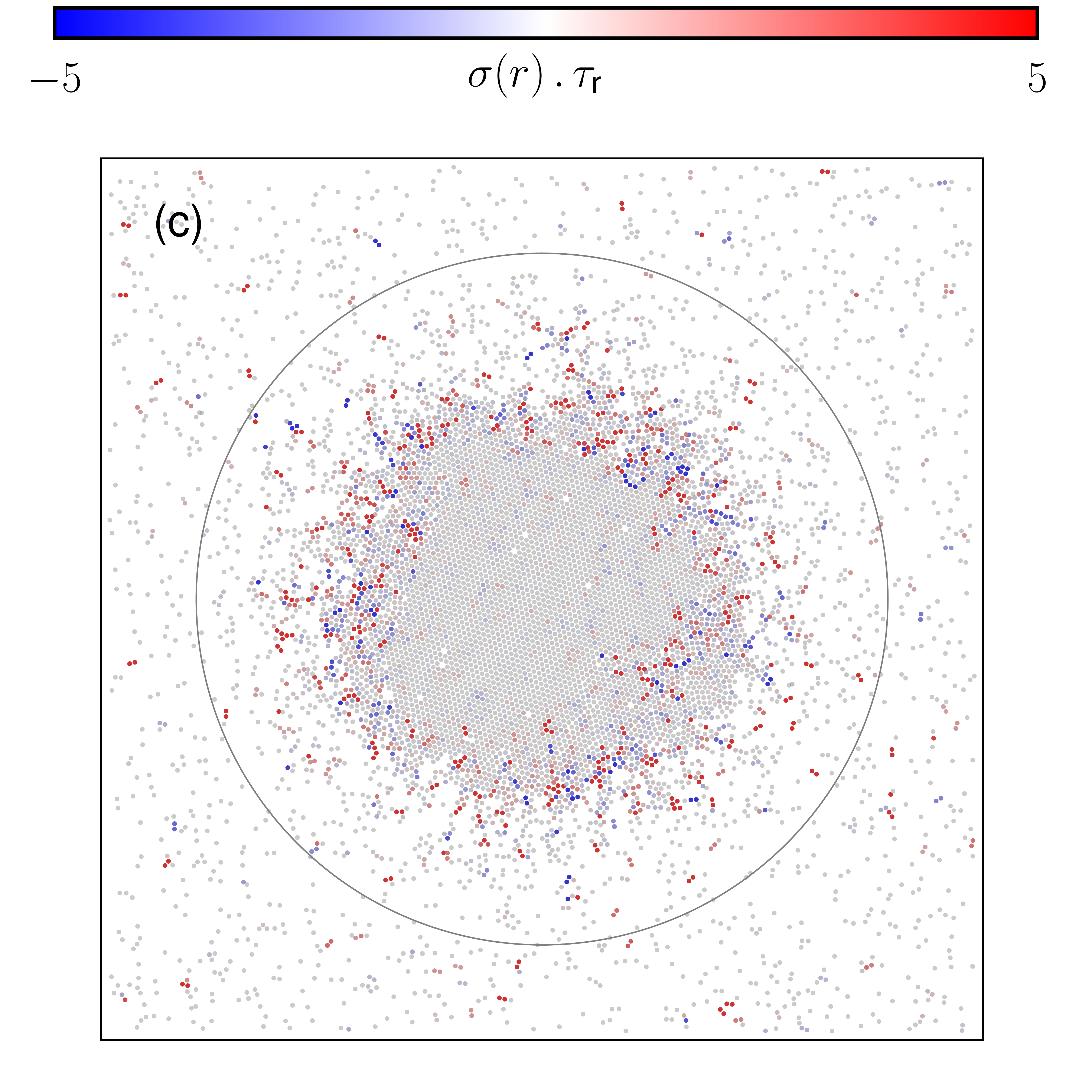}
\hfill
\includegraphics[width=\columnwidth]{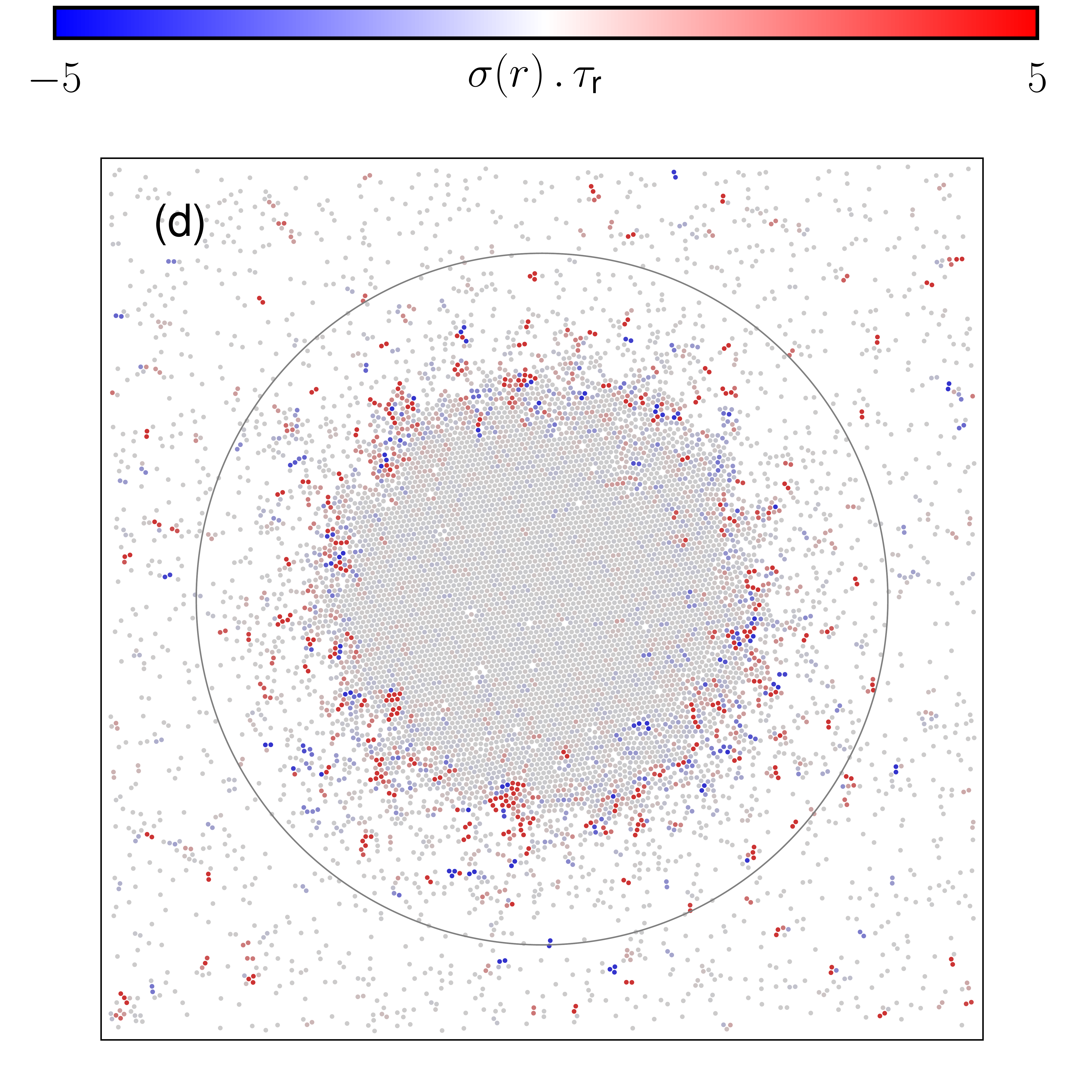}
\caption{\label{fig:EP}(Color online) Snapshots of AOUPs interacting
  via the short-range soft-core potential~\eqref{eq:pot}, and confined
  in a harmonic potential $ U (\br) = U_0 ( r / a_0 )^2 $ with finite
  range $a_0$. The range of the potential is represented by the grey
  circle. The color of each particle refers to the associated
  instantaneous value of the entropy production rate, expressed in
  units of $ 1/ \tau_\text{r} = \eps / a^2$.  We observe that
  particles form a dense compact cluster centered at the bottom of the
  harmonic trap with radial symmetry, in contact with a dilute bath of
  particles. The interface between the dense and dilute phases
  fluctuates, and the relative size of the dense phase increases with
  the number of particles from (a) to (d). Number of particles: (a)~$N
  = 5625$, (b)~$N=6750$, (c)~$N=7875$, (d)~$N=9000$. Other parameters:
  $L=150$, $D=1$, $U_0=2$, $\eps=10$, $a=1$, $\tau=10$, $a_0=60$.}
\end{figure*}

\subsection{Entropy production in MIPS}

The relationship between MIPS and equilibrium phase-separation was
initially suggested in~\cite{Tailleur:08}, using a coarse-grained
description only valid at the homogeneous level. Further discussions
on the role of gradient terms has led to the realization that this
mapping onto equilibrium is generically broken at higher order in
gradients~\cite{wittkowski2014scalar,Speck2014PRL,caballero2018strong,Solon2018:NJP,Tjhung:2018:PRX}. The
entropy production of active field theories was indeed shown to be
peaked at interfaces between coexisting phases~\cite{Nardini2016b},
a result which begs to be tested in particle models.

To do so, we first construct a map of entropy production rate starting
from the global formula~\eqref{eq:EP} for $\cS$. We consider the case
where particles interact through a pair potential:
\begin{equation}\label{eq:phi}
	\Phi = \f{1}{2} \sum_{ i, j = 1 }^N V ( \br_i - \br_j ) ,
\end{equation}
where $ V(0) = 0 $ to avoid self-interaction. To rewrite~\eqref{eq:EP}
as a sum over particle contributions, we first note that
\begin{equation}
	\begin{aligned}
		& \avg{ ( \bp_i \cdot \nabla_i )^3 \Phi }
		\\
		& \quad =\frac 1 2 \sum_{ n,p,q,i,j = 1 }^N \avgg{ ( \bp_n \cdot \nabla_n ) ( \bp_p \cdot \nabla_p ) ( \bp_q \cdot \nabla_q ) V ( \br_i - \br_j ) } .
	\end{aligned}
\end{equation}
Using that
\begin{equation}
  \sum_k ( \bp_k \cdot \nabla_k ) V ( \br_i - \br_j ) = (\bp_i-\bp_j) \cdot \nabla_i V(\br_i-\br_j)
\end{equation}
the entropy production rate can be rewritten as
\begin{equation}\label{eq:EP2}
	\cS =  \sum_{ i = 1 }^N \sigma_i\;,
\end{equation}
where $\sigma_i$ is given by
\begin{equation}\label{eq:sigmai}
  \sigma_i=\avg{ \f{ \tau^2 }{4D}\sum_{ j = 1 }^N \big[ \pnt{ \bp_i - \bp_j } \cdot \nabla_i\big]^3 V ( \br_i - \br_j ) }\;.
\end{equation}
Equations~\eqref{eq:EP2} and~\eqref{eq:sigmai} split the total entropy
production rate into a sum of particle contributions. It is thus
tempting to refer to $\sigma_i$ as the entropy production rate of
particle $i$. Note, however, that we have proven neither that
$\sigma_i$ is positive nor that the decomposition~\eqref{eq:sigmai} is
unique. This interpretation thus has to be taken with a pinch of salt.

Our aim is to compute the local rate of entropy production associated
with a phase-separated set of AOUPs, and in particular investigate the
correlation between the structure of the density field and entropy
production. We consider $N$ particles in a $2d$ box, interacting via a
short-range soft-core potential of the form
\begin{equation}\label{eq:pot}
  V (\br) = \eps \exp\brt{ - \f{ 1 }{  ( r /a )^2 - 1 } }\;,
\end{equation}
for $r\leq a$. As discussed in Sec.~\ref{sec:MIPSPFAPs}, such pairwise
repulsive forces can lead to phase separation at high enough
persistence: a macroscopic cluster of particles then forms and slowly
diffuses in the system. This slow diffusion makes the comparison
between the average density and the entropy production rate profiles
numerically difficult and we thus pin the center of the cluster by
applying a truncated harmonic trap of the form $ U (\br) = U_0
(r/a_0)^2$ for $r\leq a_0$. Note that the harmonic trap does not
directly contribute to the entropy production since it is harmonic.

\begin{figure*}
\includegraphics[width=\columnwidth]{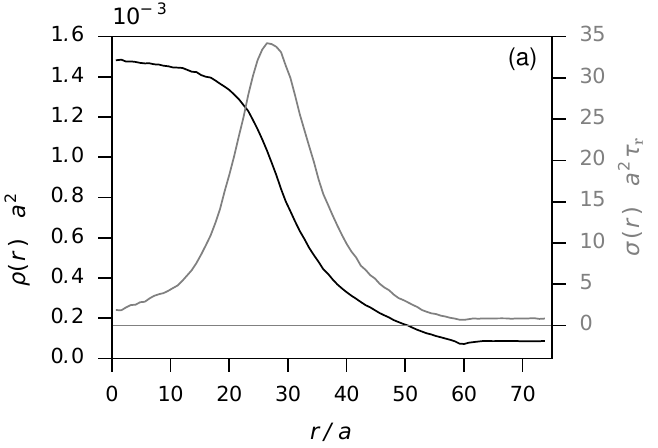}
\hfill
\includegraphics[width=\columnwidth]{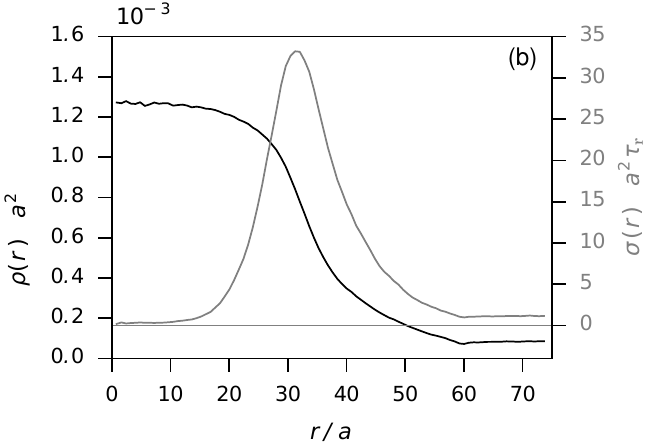}
\vskip.5cm
\includegraphics[width=\columnwidth]{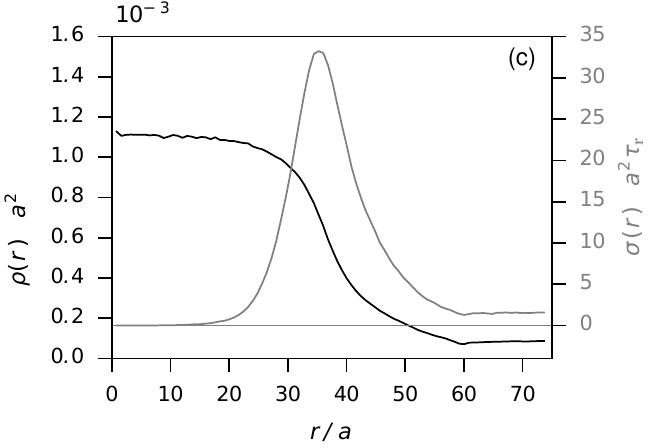}
\hfill
\includegraphics[width=\columnwidth]{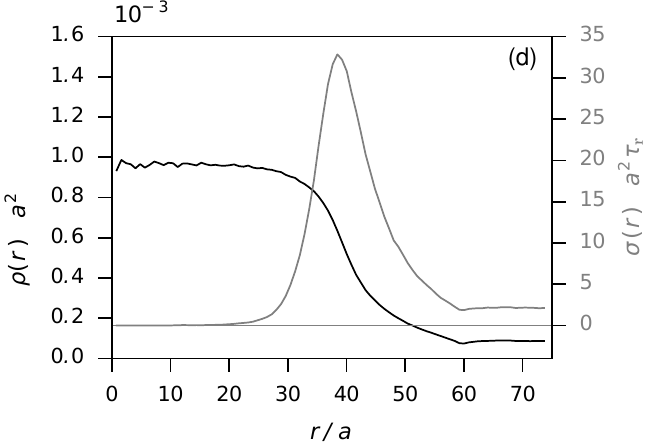}
\caption{\label{fig:EP_profile}(Color online) Density and local
  entropy production rate profiles as functions of the distance from
  the centre of the harmonic trap. Parameters are the same as in
  Fig.~\ref{fig:EP}.
}
\end{figure*}

As show in Fig.~\ref{fig:EP} for different particle densities, a
macroscopic cluster of particles indeed localizes in the center of the
system and coexists with a surrounding dilute gas phase. Increasing
the density mostly results in a shift of the boundary between the two
phases as the cluster grows. Using Eq.~\eqref{eq:sigmai}, the
contribution of each particle to the entropy production can be
measured: particles in the bulk of each phase have negligible
contributions whereas those localized at the boundary of the cluster
exhibit large values, both positive and negative, of $\sigma_i$. To
compute the net contribution to the entropy production, we construct a
density $\sigma(r)=\overline{\sum_{i=1}^N \sigma_i \delta(r-r_i)}$ of
entropy production rate, where the overbar represents a binning
procedure and a time average in the steady-state. The resulting
profiles are shown in Fig~\ref{fig:EP_profile}, where they are
compared to the density profiles.

As hinted from Fig.~\ref{fig:EP}, $\sigma(r)$ is indeed much smaller
in bulk phases than at the boundary of the dense cluster, which
highlights that the breakdown of time-reversal symmetry in MIPS is
dominated by interface physics. Novelty with respect to a \textit{bona
  fide} equilibrium phase separation will thus mostly operate at the
interface between dilute and dense phases, which echoes recent results
on reversed Ostwald ripening and bubbly phase-separations in active
systems~\cite{Tjhung:2018:PRX,caballero2018bulk}. The almost
negligible value of $\sigma_i$ in both bulk phases can be understood by
inspection of~\eqref{eq:sigmai}: $\sigma_i$ vanishes in the dense
phase since the relative velocities between the particles vanish, and
it reaches a small plateau value in the dilute phase, where
collisions are scarce. On the contrary, $\sigma(r)$ reaches a maximum
value at the interface where fast particles coming from the dilute
phase collide onto slow ones arrested in the crowded cluster.

Note that our microscopic analysis corroborates the phenomenological
coarse-grained approach of~\cite{Nardini2016b}. This validates
\textit{a posteriori} the idea that, at the coarse-grained scale, the
difference between MIPS and an equilibrium phase-separation lies in
non-conservative gradient terms contributing mostly at the
interface. Finally, the differences between all microscopic models
exhibiting MIPS-like behaviours, whether on- or off-lattice,
interacting via quorum-sensing or pairwise forces, should be mostly
apparent, at the coarse-grained scale, in the different types of
gradient terms they will generate.

\subsection{Symmetry of time correlations}
While entropy production rate provides a formal estimate of the breakdown
of time-reversal symmetry, its practical significance may be hard to
grasp. It is furthermore not easily accessible experimentally, and its
measurement has thus been carried out mostly on low-dimensional
systems~\cite{Andrieux2007entropy,roldan2018arrow}. A more pragmatic
measurement of irreversibility can be inferred from the asymmetry of
time-correlations of physical observables. For systems in equilibrium,
time-reversibility indeed implies that time-correlations are
symmetric. Consider first an overdamped equilibrium system whose
configurations are specified by the positions $\br =\{\br_i\}$ of the
particles. For any two observables $A(\br)$ and $B(\br)$, time
reversibility implies that
\begin{equation}\label{eq:symm-AB-BA-eq-even}
  \langle A(0)B(t)\rangle -\langle A(t)B(0)\rangle=0
\end{equation}
where the average is defined as
\begin{equation}\label{eq:defaverage}
  \langle A(0)B(t)\rangle\equiv \int d \br P_s(\br) A(\br) e^{t
    \cL^\dagger} B(\br)\;.
\end{equation}
In~\eqref{eq:defaverage}, $\cL$ is the Fokker-Planck operator and
$P_s(\br)$ the corresponding steady-state distribution.

Equation (\ref{eq:symm-AB-BA-eq-even}) can be generalized to systems
with odd variables under time-reversal ($\bp =\{\bp_i\}$), such as,
for instance, underdamped dynamics. As discussed in Appendix~\ref{sec:TTIandTRS},
considering observables $A=A(\br,\bp)$ and $B=B(\br,\bp)$,
time-reversibility now implies that
\begin{equation}\label{eq:symm-AB-BA-eq-def}
  \Delta_{A,B}(t)\equiv\langle A(0)B(t)\rangle -\langle (\Pi A)(t) (\Pi B)(0)\rangle=0
\end{equation}
where $\Pi A(\br,\bp)=A(\br,-\bp)$ and analogously for $B$. In
Eq.~\eqref{eq:symm-AB-BA-eq-def}, the average is defined as
in~\eqref{eq:defaverage}, albeit with the Fokker-Planck operator $\cL$
now acting on function $P_{\rm s}(\br,\bp)$ defined on the full phase
space. The vanishing of $\Delta_{{A,B}}$ in
Eq.~\eqref{eq:symm-AB-BA-eq-def} simply reflects that trajectories
$\br(s),\bp(s)$ of length $t$ are as likely as the reversed ones
$\br(t-s),-\bp(t-s)$ obtained by flipping the momenta.

AOUPs possess an effective equilibrium regime, characterised by a
vanishing entropy production rate for small $\tau$, and we thus expect
(\ref{eq:symm-AB-BA-eq-def}) to be valid in such a limit. Let us first
show how the computation of $\Delta_{A,B}(t)$ and of the entropy
production rate bare some similarities. Using the path-integral
methods presented in section~\ref{sec:sig}, the correlator $\langle
\Pi A(t)\Pi B(0)\rangle$ can be written as
\begin{eqnarray}
  \langle \Pi A( t)\Pi B(0)\rangle &=& \int {\cal D}[\bq,\bp]    P_{\rm s}[\bq(0),\bp(0)] \delta(\dot{\bq}-\bp)\nonumber\\
  && e^{-\cA[ \bq,\bp]}  \Pi A( t)\Pi B(0)
\end{eqnarray}
where $\cA$ is the dynamical action defined
in~\eqref{eq:defaction}. Changing variable from $(\bq(s),\bp(s))$ to
$(\bq^\R(s),\bp^\R(s))=(\bq(t-s),-\bp(t-s))$,
this can be recast into
\begin{eqnarray}
  \langle \Pi A( t) \Pi B(0)\rangle &=& \int {\cal D}[\bq^\R,\bp^\R] \delta(\dot{\bq}^\R-\bp^\R)P_{\rm s}[\bq^\R(t),-\bp^\R(t)]\nonumber\\
  & & \hspace{-2.25cm} e^{-\cA [\bq^\R,\bp^\R]+\delta\cA[\bq^\R,\bp^\R]}
   A(\bq^\R(0),\bp^\R(0)) B(\bq^\R(t),\bp^\R(t))
\end{eqnarray}
where we have used that $\delta\cA[\bq^\R,\bp^\R]=
\cA[\bq^\R,\bp^\R]-\cA[\bq,\bp]$. Finally, using that
\begin{equation*}
  P_{\rm s}[\bq^\R(t),-\bp^\R(t)]= P_{\rm s}[\bq^\R(0),\bp^\R(0)]
  e^{\log\frac{P_{\rm s}[\bq^\R(t),-\bp^\R(t)]}{P_{\rm s}[\bq^\R(0),\bp^\R(0)]}}
\end{equation*}
and dropping the superscript ${}^\R$ on the dummy variables $\bq^\R,\bp^\R$
leads to:
\begin{equation}\label{eq:symm-AB-BA-eq}
  \Delta_{A,B}(t)= \left\langle A(0) B(t) \left(1-e^{\delta {\cal A}[\bq,\bp] + \log \frac{{P_{\rm s}[\bq(t),-\bp(t)]}}{P_{\rm s}[\bq(0),\bp(0)]}}\right)\right\rangle
\end{equation}
The connection to entropy production then comes from the fact that
\begin{equation}\label{eq:deltaandsigma}
  \cS = - \lim_{t\to\infty} \frac 1 t \left(\delta {\cal A} + \log \frac{{P_{\rm s}[\bq(t),-\bp(t)]}}{P_{\rm s}[\bq(0),\bp(0)]}\right)
\end{equation}
We have shown in section~\ref{sec:sig} that the entropy production
\textit{rate} scales as $\cS \propto \tau^2$ when $\tau\to
0$. Equation~\eqref{eq:symm-AB-BA-eq} instead involves the total
entropy production during a time $t$. Approximating the
latter by $t \cS$ would imply that $\Delta_{A,B}\sim (1-e^{-\cS
  t})\to 0$ as long as $t\ll \tau^{-2}$. This approximation is,
however, uncontrolled, and one has to deal, in practice, with the
boundary terms that vanish in the computation of $\cS$.

In this section, we thus follow an alternative path, based on an
operator formalism, that allows us to treat the small $t$ limit
exactly but offers bounds on the decay of $\Delta_{A,B}$ with $\tau$
which are weaker than those derived for $\cS$. In the limit $\tau\to
0$, the generator of the AOUPs process, i.e. the adjoint of the
Fokker-Planck operator, is time-reversal symmetric. As $\tau$
increases, it develops a non-symmetric part, which can be computed
using our expansion of the steady-state distribution for small
$\tau$. As detailed in Appendix \ref{app:time-correlations} the
leading order of the antisymmetric part of the generator is given by
$\tau \mathcal{L}_A^{\dagger}$, where
\begin{equation}\label{eq:LHP-AOUPS-Antysymm}
  \mathcal{L}^{\dagger}_A (\bq,\bar{\bp})=
  \Big\{(\nabla_i\nabla^2\Phi)
  - \, [(\bar{\bp}_j\cdot \nabla_j)^2\nabla_i\Phi]\Big\}\cdot
  \frac{\partial}{\partial \bar{\bp}_i}\,.
\end{equation}
and we work again with the rescaled variable $\bar \bp= \sqrt\tau
\bp$. Using perturbation theory~\cite{Baiesi:13}, we show in Appendix
\ref{app:time-correlations} that
\begin{equation}\label{eq:Delta-AB-explicit}
  \begin{aligned}
    &\Delta_{A,B}(t= \tau u)=\tau^{3/2}\int_0^u du_1\, \int d\bq\,d\bar{\bp}
    \,
    P_{s}(\bq,\bar{\bp}) A(\bq,\bar{\bp})\\
    &e^{u_1\sqrt{\tau}\mathcal{L}^{\dagger}(\bq,\bar{\bp})}
    \mathcal{L}_A^{\dagger}(\bq,\bar{\bp})
    e^{(u-u_1)\sqrt{\tau}\mathcal{L}^{\dagger}(\bq,\bar{\bp})} B(\bq,\bar{\bp})+ O(\tau^2)\,
  \end{aligned}
\end{equation}
Noting that
$\sqrt{\tau}\mathcal{L}^{\dagger}\sim\mathcal{O}(\tau^0)$, we conclude that,
for short times,
\begin{equation}\label{eq:Delta-AB}
  \begin{aligned}
    \Delta_{A,B}(t=\tau u)\sim \mathcal{O}(\tau^{3/2})\,\qquad \textrm{for }\qquad u\sim \mathcal{O}(\tau^0)\,.
  \end{aligned}
\end{equation}

A stronger result than (\ref{eq:Delta-AB}) actually holds when at least one of the two observables, say $B$, depends only  on positions: $B=B(\bq)$. In this case, indeed,
\begin{equation}
  \begin{aligned}\label{eq:deltaAB-appr}
    e^{(u-u_1)\sqrt{\tau}\mathcal{L}^{\dagger}(\bq,\bar{\bp})} B(\bq)
    &=e^{(u-u_1)\sqrt{\tau}\mathcal{L}_0^{\dagger}(\bq,\bar{\bp})} B(\bq) + \mathcal{O(\sqrt{\tau})}\\
    &=B(\bq) + \mathcal{O(\sqrt{\tau})}
  \end{aligned}
\end{equation}
where the first equality comes from expanding $\cL$ and retaining only
its leading order in $\tau$, $\cL_0$, which is given by:
\begin{equation}
  \begin{aligned}
    \sqrt{\tau}\cL_0 \equiv  \f{\p}{ \p p_{i\alpha} } p_{i\alpha}
    + D  \f{\p^2}{ \p p_{i\alpha}^2 } .
  \end{aligned}
\end{equation}
The second equality then follows by noticing that $\cL_0$ acts only on
momenta. Inserting (\ref{eq:deltaAB-appr}) in
Eq.~(\ref{eq:Delta-AB-explicit}) leads to
\begin{equation}
	\begin{aligned}
	\Delta_{A,B}(t=\tau u) = \mathcal{O}(\tau^2)\,\qquad u\sim \mathcal{O}(\tau^0)\,.
	\end{aligned}
\end{equation}
As announced earlier, the effective time-reversal symmetry of AOUPs
dynamics for small $\tau$ thus leads to a corresponding symmetry for
two-time correlation functions. The convergence as $\tau\to 0$ is
faster for position-dependent observables than in the full phase-space
$\br,\bar \bp$, which is consistent with the fact that momenta-like
variables have to be handled with care in the context of
TRS~\cite{mandal2017entropy,caprini2018comment,caprini2019entropy}.


\section{Effective equilibrium: Linear response and fluctuation-dissipation relations}\label{sec:fdr}

A most celebrated way of probing the nonequilibrium nature of the
dynamics lies in investigating the connection between fluctuations and
responses.  For passive inertial Brownian particles~\eqref{eq:K}, a
perturbation of the potential $ \Phi \to \Phi - h B $ amounts to
modifying the dynamics into
\begin{equation}
	\tau \dot \bp_i = - \bp_i - \nabla_j \Phi + h \nabla_j B + \sqrt{2D} \bet_j .
\end{equation}
We define the response function $R$ which measures the effect of the perturbation on the average value of an arbitrary observable $A$ as
\begin{equation}
	R (t, s) = \left. \f{ \delta \avg{ A (t) } }{ \delta h (s) }\right|_{h=0} .
\end{equation}
Causality enforces that it vanishes when $t<s$, and time-translation
invariance in the steady-state means that it only depends on the time
difference $t-s$. In equilibrium, fluctuations and response are
related by the celebrated Fluctuation Dissipation Theorem (FDT).  The
latter states that the response function is related to the decay rate
of the (unperturbed) correlations between $A$ and $B$:
\begin{equation}\label{eq:FDTEQ}
  R (t, s) = - \f{1}{D} \f{\dd}{\dd t} \avg{ A (t) B (s) } .
\end{equation}
Our aim is to determine to which extent one can obtain a
fluctuation-dissipation relation (FDR) analogous to~\eqref{eq:FDTEQ}
for the active dynamics~\eqref{eq:dyn}, a topic which has attracted
interest
recently~\cite{szamel2017evaluating,caprini2018linear,dal2019linear,DalCengio2020,maes2020response}. These
works are based on the explicit expression of the steady-state
distribution of AOUPs. Here, instead, we follow~\cite{Fodor2016PRL}
and proceed at a dynamical level to derive generic relations between
response and correlation functions, which do not require any knowledge
of the steady-state. We then study the response to perturbing the
strength of self-propulsion in the effective equilibrium regime.


\subsection{Perturbing with an arbitrary potential}
\label{subsec:perturbing-potential}

Under the perturbation $ \Phi \to \Phi - h B(\{\br\}) $, the dynamics~\eqref{eq:dyn} changes as
\begin{equation}
	\dot \br_i = - \nabla_i \Phi + h \nabla_i B + \bv_i \;.
\end{equation}
To compute the response function of an observable $A(\{\br\})$, we
write the dynamics for $\bp_i=\dot \br_i$  as
\begin{equation}
\tau \dot \bp_i = - \bp_i - \left(1 + \tau \frac{\rm d}{{\rm d}t} \right) (\nabla_i
\Phi-h \nabla_i B)  + \sqrt{2D} \bet_i
.
\end{equation}
Using the formalism of section~\ref{sec:sig}, the dynamical action becomes
\begin{equation}
	\cA = \f{1}{4D} \int_0^t \bigg | \pnt{ 1 + \tau \f{\dd}{\dd u} } \pnt{ \bp_i + \nabla_i \Phi - h \nabla_i B } \bigg |^2 \dd u .
\end{equation}
The probability of a trajectory is given by
$\cP(\{\br(t),\bp(t)\})\sim \exp[-\cA[\{\br(t),\bp(t)\}]]$, so that $\delta \cP = - \delta \cA . \cP$, leading to a response function
given by
\begin{equation}
	R (t,s) = - \avg{ A (t) \left. \f{ \delta \cA }{ \delta h(s) } \right|_{h=0} } \;.
\end{equation}
Note that it is sufficient to compute $ \delta \cA $ to first order in $h$:
\begin{equation}
	\delta \cA = - \f{1}{2D} \int_0^t h \nabla_i B \cdot \pnt{ 1 - \tau^2 \f{\dd^2}{\dd u^2} } ( \bp_i + \nabla_i \Phi ) \dd u + \cO (h^2)
\end{equation}
to determine the response function. We now follow standard procedures~\cite{Cugliandolo:11, Baiesi:13} and we consider the difference $R(t,s)-R(-t,-s)$ in the steady-state where time-translation invariance means that $R(t,s)=R(t-s)$. The difference of response functions reads
\begin{equation}\label{eq:resp-detailed}
	\begin{split}
	R(t-s)&-R(s-t)=\\
	&\frac{1}{2D}\left\langle A(t)\left[\nabla_i B\left(1-\tau^2\frac{\dd^2}{\dd s^2}\right)(\bp_i+\nabla_i\Phi)\right](s)\right\rangle\\
	-&\frac{1}{2D}\left\langle A(-t)\left[\nabla_i B\left(1-\tau^2\frac{\dd^2}{\dd s^2}\right)(\bp_i+\nabla_i\Phi)\right](-s)\right\rangle
	\end{split}
\end{equation}
We now use that, due to time-translation invariance in the
steady-state, for any two observables $A$ and $C$ we have
$\langle A(-t)C(-s)\rangle=\langle A(s) C(t)\rangle$. This leads to
\begin{equation}\label{eq:resp-detailed-2}
	\begin{split}
	  R(t-s)-&R(s-t)=\frac{1}{2D}\langle A(t)\left[\nabla_i B\left(1-\tau^2\frac{\dd^2}{\dd s^2}\right)\bp_i\right](s)\rangle\\
          &-\frac{1}{2D}\langle A(s)\left[\nabla_i B\left(1-\tau^2\frac{\dd^2}{\dd t^2}\right)\bp_i\right](t)\rangle\\
	&+\frac{1}{2D}\langle A(t)\left[\nabla_i B\left(1-\tau^2\frac{\dd^2}{\dd s^2}\right)\nabla_i\Phi\right](s)\rangle\\
	&-\frac{1}{2D}\langle A(s)\left[\nabla_i B\left(1-\tau^2\frac{\dd^2}{\dd s^2}\right)\nabla_i\Phi\right](t)\rangle
	\end{split}
      \end{equation}
      \if{
Noting that $\bp_i(t)=\frac{\dd}{\dd t}\br_i$, the second line
of~\eqref{eq:resp-detailed-2} can be rewritten as
\begin{equation}
  \langle A(s)\left[\nabla_i B\left(1-\tau^2\frac{\dd^2}{\dd t^2}\right)\bp_i\right](t)\rangle = \frac{\dd}{\dd t} f(s-t)
\end{equation}
where we have introduced $f(s-t)= \langle A(s)[\nabla_i
B(1-\tau^2\frac{\dd^2}{\dd t^2})\br_i](t)\rangle$.}\fi
Using causality, we have that, for $s<t$, $R(s-t)=0$.
\if{This finally leads to
\begin{equation}\label{eq:resp-detailed-3}
	\begin{split}
	R(t-s)=&\frac{1}{2D} \frac{\dd}{\dd s} f(t-s)-\frac{1}{2D} \frac{\dd}{\dd t} f(s-t)\\
	&+\frac{1}{2D}\langle A(t)\left[\nabla_i B\left(1-\tau^2\frac{\dd^2}{\dd s^2}\right)\nabla_i\Phi\right](s)\rangle\\
	&-\frac{1}{2D}\langle  A(s)\left[\nabla_i B\left(1-\tau^2\frac{\dd^2}{\dd s^2}\right)\nabla_i\Phi\right](t)\rangle
	\end{split}
\end{equation}}\fi
Furthermore, in the effective equilibrium regime, the symmetry under
a time reversal enforces that the last two lines in~\eqref{eq:resp-detailed-2}
cancel out, whereas the first two add up. All in all,
the response finally reads
\begin{equation}\label{eq:resp_fin}
	R (t-s) = - \f{1}{D} \f{\dd}{\dd t} \avg{ A (t) B (s) } - \f{\tau^2}{D} \avg{ A(t) ( \ddot \bp_i \cdot \nabla_i ) B (s) } .
\end{equation}
By contrast to thermal equilibrium, the response is not the time
derivative of a given function in general. There are exceptions, for
instance when $B=\sum_i f_i \br_i$, where $f_i$ are a set of constant
forces exerted on the particles, as considered in~\cite{Fodor2016PRL}.

The two approximate treatments presented previously, based on either UCNA or the Fox theory, lead to Markovian dynamics for the particle positions. Therefore, we can use previous results which predict the form of the response in terms of the stationary distribution~\cite{agarwal1972,Prost:09}:
\begin{equation}
	R (t, s) = - \f{\dd}{\dd t} \avg{ A (t) \left. \f{ \p \ln P_\sS }{ \p h } \right|_{h=0} (s) } ,
\end{equation}
where $P_\sS$ denotes the stationary distribution under the perturbed potential $ \Phi - h B $. From~\eqref{eq:Ps_ucna},
we linearize the distribution around the unperturbed state as
\begin{equation}
	\begin{aligned}
		D \ln P_\sS &= - \Phi - \f{\tau}{2} ( \nabla_i \Phi )^2 + D \ln \abs{ \det \M }
		\\
		& \quad + h \pnt{ B + \tau \nabla_i \Phi \cdot \nabla_i B - \tau D \nabla_i^2 B } + \cO (h^2) ,
	\end{aligned}
\end{equation}
yielding
\begin{equation}\label{eq:LRFOX}
	R (t) = - \f{1}{D} \f{\dd}{\dd t} \avg{ A(t) \pnt{ B + \tau \nabla_i \Phi \cdot \nabla_i B - \tau D \nabla_i^2 B } (s) } .
\end{equation}
The result~\eqref{eq:LRFOX}, which stems from the Fox theory and UCNA, differs from~\eqref{eq:resp_fin}. This shows that
UCNA and the Fox theory have to be used with care when studying
dynamical observables, even to first order in $\tau$.


\subsection{Perturbing the amplitude of fluctuations}
Let us now consider a perturbation $D\to D(1 + h \Theta(t))$ in
Eq.~(\ref{eq:OU}), which can be seen as a change in the typical
self-propulsion speed $\sqrt{D/\tau}$ at fixed persistence time. Such
perturbation is particularly interesting given the recent experimental
development of self-propelled particles, both of
synthetic~\cite{Speck:13,Palacci:13,kummel2013circular,palacci2014light}
and
biological~\cite{walter2007light,vizsnyiczai2017light,frangipane2018dynamic,arlt2018painting,arlt2019dynamics}
natures, whose self-propulsion speed can be controlled by external
light sources.

Let us first recall what happens in equilibrium when perturbing the
temperature, an operation we denote by $T\to T(1 + h \Theta(t))$. Both
in the over-damped and in the under-damped cases, the effect of such a
perturbation on an arbitrary observable $A$ can be written as~\cite{agarwal1972,baiesi2014thermal}
\begin{equation}
	\begin{aligned}\label{eq:pert-T-equilibrium}
\langle \delta A(t)\rangle
= \frac{h}{T} \, \langle H(0)[A(0)-A(t)]\rangle \,,
	\end{aligned}
\end{equation}
where $H$ is the energy of the system, we set the Boltzmann constant
to unity, and $\langle \delta A(t)\rangle = \langle A(t) \rangle_{h} -
\langle A(0)\rangle $. Here, the average $\langle \cdot \rangle_{h}$
is computed in the presence of the perturbation.

Let us now consider the case of AOUPs. In
Appendix~\ref{app:pert_temp}, we show that a similar relation can be
derived for times $t \sim \tau$:
\begin{equation}
  \begin{aligned}\label{eq:entr-flux-AOU-new-time}
    \langle \delta A(t= s \tau)\rangle
      =\frac{h}{D}\Big\langle H_{\rm eff}(0) \, [A(0)-A(t=s \tau)] \Big\rangle
    +O(\tau^{3/2})
  \end{aligned}
\end{equation}
where
\begin{equation}
  \begin{aligned}\label{eq:effective-H}
    H_{\rm  eff}&=H_0+\tau  H_1 \\
    H_0&=\Phi + \frac{\bar{\bp}_i^2}{2}\\
    H_1&=  \frac{1}{2}\left[\pnt{ \nabla_i \Phi }^2 + \pnt{ \bar{\bp}_i \cdot \nabla_i }^2 \Phi - 3 D \nabla_i^2 \Phi  \right] \,.
  \end{aligned}
\end{equation}
The above result is the generalisation to AOUPs of the equilibrium
result (\ref{eq:pert-T-equilibrium}). Interestingly, $H_{\rm eff}$ is
the effective Hamiltonian one would infer from the logarithm of the
stationary measure~\eqref{eq:Pss} up to order $\mathcal{O}(\tau)$. At
odds with equilibrium systems where eq. (\ref{eq:pert-T-equilibrium})
is valid for all times, Eq. (\ref{eq:entr-flux-AOU-new-time}) is only
derived here for short trajectories.


\section{Dynamics of collective modes }\label{sec:coll_modes}

\if{In this section we study the dynamics of the fluctuating microscopic density and velocity modes. They are defined in terms of the particles' positions and velocities by
\begin{equation}
	\rho (\br, t) = \sum_{i=1}^N \delta (\br - \br_i(t) ) , \quad g_\alpha (\br, t) = \sum_{i=1}^N p_{i\alpha} \delta ( \br - \br_i(t) ) .
\end{equation}
Note that there is less information contained in  the density field $\rho$ and the velocity field $g_\alpha$ than in the individual positions $\br_i$ and velocities $p_{i\alpha}$.
Neither $\rho$ nor $g_\alpha$ are hydrodynamic fields in the sense that they are not built from a mesoscopic coarse-graining procedure. They are still fully microscopic (and as such highly singular) objects.
For particles evolving according to overdamped Langevin equations, Kawasaki and Dean have shown the steps towards an exact partial differential stochastic equation for the local density field $\rho$~\cite{Kawasaki:94, Dean:96}. A similar procedure, now involving both $\rho$ and $g_\alpha$, was later achieved for particles evolving according to underdamped Langevin dynamics~\cite{Nakamura:09}. Bridging exact equations for the microscopic $\rho$ and $g_\alpha$ to their hydrodynamic counterparts requires to resort to a vaerity of approximations even if the equations for $\rho$ and $g_\alpha$ appear to be closed. Closure at the hydrodynamic level has been discussed by several authors up to very recently~\cite{Das-Mazenko:1986,Dufty:1993,Archer:06, Archer:09, Das:13, Demery:15} (see also \cite{Katie-Kranthi:2017, epstein2019statistical} for a more recent discussion within the framework of active matter). The main interest of having exact equations at our disposal is to generate BBGKY-like hierarchies or even to implement approximation methods with no analog within the Lagrangian particle-based formulation. Before we briefly explore possible uses of evolution equations for $\rho$ and $g_\alpha$ we explain how they can be derived.
}\fi


In this section we study the dynamics of the fluctuating microscopic density and velocity modes. They are defined in terms of the particles' positions and velocities by
\begin{equation}
	\rho (\br, t) = \sum_{i=1}^N \delta (\br - \br_i(t) ) , \quad g_\alpha (\br, t) = \sum_{i=1}^N p_{i\alpha} \delta ( \br - \br_i(t) ) \;,
\end{equation}
which are sometimes referred to as empirical measures.  Neither $\rho$
nor $g_\alpha$ are hydrodynamic fields in the sense that they are not
built from a mesoscopic coarse-graining procedure. They are still
fully microscopic (and as such highly singular) objects and one can
thus hope to obtain exact evolution equations for them.

For particles evolving according to overdamped Langevin equations,
Kawasaki and Dean have shown the steps towards an exact partial
differential stochastic equation for the local density field
$\rho$~\cite{Kawasaki:94, Dean:96}. Extracting physical information at
hydrodynamic level out of the Dean-Kawasaki equation often requires,
however, a further crucial approximation: to replace $\rho$ with a
smooth field $\rho^H$, the hydrodynamic field, and assume that
$\rho^H$ still solves the same Dean-Kawasaki equation. Indeed, it is
an implicit assumption of most numerical and analytical approaches
that one looks only for solutions with sufficient degree of
regularity, which $\rho^H$ is assumed to have while $\rho$ clearly has
not.

We are aware of only one case where it is possible to prove that the smoothing procedure described above is harmless, in the sense that the evolution of $\rho^H$ obeying the Dean-Kawasaki equation approximates precisely the evolution of $\rho$. It happens for overdamped particles interacting with a weak potential, that scales as $1/N$. This is a relevant model for either systems with long-range interactions~\cite{campa2009statistical} such as interactions mediated by a surrounding low-Reynolds number fluid~\cite{stenhammar2017role} or for very soft bodies such as polymers~\cite{doi1988theory}. In this case, comparing rigorous mathematical results~\cite{dawsont1987large} with those obtained from the Dean-Kawasaki equation allow to conclude that $\rho^H$ captures not only the average evolution of $\rho$ but even its fluctuations up to those whose probability is exponentially small in $N$~\cite{nardiniperturbative2016}. For systems without weak interactions, it remains an outstanding open question the degree of approximation that is involved in passing from the empirical measure $\rho$ to the smooth hydrodyamic field $\rho^H$.

In the case of systems with underdamped dynamics, a similar procedure to the one leading to the Dean-Kawasaki equation, but now involving both $\rho$ and $g_\alpha$, was later achieved in~\cite{Nakamura:09}. In this case, passing to the hydrodynamic limit would require further assumptions with respect to the case of overdamped particles, even if the equations for $\rho$ and $g_\alpha$ appear to be closed. The main point is that, while some of the steps needed to derive closed equations for $\rho$ and $g_\alpha$ are exactly true when working at the level of the empirical fields $\rho$ and $g_\alpha$, it is unclear whether they remain valid in the smoothing required to pass from them to their hydrodynamics counterparts.
Taking a different approach, closure at the hydrodynamic level has been discussed by several authors up to very recently~\cite{Das-Mazenko:1986,Dufty:1993,Archer:06, Archer:09, Das:13, Demery:15} (see also \cite{Katie-Kranthi:2017, epstein2019statistical} for a more recent discussion within the framework of active matter).

We will not consider the issue due to passing from empirical measures to hydrodynamic fields further in what follows and concentrate instead on deriving the exact equations  for $\rho$ and $g_{\alpha}$ for AOUPs. These could be employed to pass to the hydrodynamic level of description and generate BBGKY-like hierarchies or implement approximation methods with no analog within the Lagrangian particle-based formulation, but their development goes beyond the scope of the present paper. We limit ourselves, at the end of the Section, to sketch future research directions along these lines.


\subsection{Fluctuating hydrodynamics}

We first derive the set of equations ruling the coupled dynamics of $\rho$ and $\bg$. The time derivative of the local density of course takes the form of a continuity equation
\begin{equation}\label{continuity}
	\p_t \rho (\br, t) = - \p_\alpha \sum_{i=1}^N \dot r_{i\alpha} \delta ( \br - \br_i(t) ) = - \p_\alpha g_\alpha (\br, t),
\end{equation}
thus expressing the conservation of the number of particles. The time derivative of the current density $g_\alpha$ is given by
\begin{equation}\label{eq:g}
	\p_t g_\alpha (\br, t) = - \p_\beta \sum_{i=1}^N p_{i\alpha} p_{i\beta} \delta ( \br - \br_i(t) ) + \sum_{i=1}^N \dot p_{i\alpha} \delta ( \br - \br_i(t) ) .
\end{equation}
To proceed further, we assume that the particles interact through pairwise interactions only as given in~\eqref{eq:phi}. When substituting the microscopic dynamics~\eqref{eq:dyn_active_g} in the expression for the time derivative of $\bg$, a term of the following form appears:
\begin{equation}
	\begin{aligned}
		&\sum_{ i,j = 1}^N p_{j\beta} \p_{i\alpha} \p_{j\beta} \Phi \delta (\br - \br_i(t) )
		\\
		& \, = \sum_{ i,j,k = 1 }^N p_{j\beta} \p_{i\alpha} \p_{j\beta} V ( \br_i - \br_k ) \delta ( \br - \br_i(t) )
		\\
		& \, = \sum_{ i,k = 1 }^N \left( p_{i\beta} \p_{i\beta} + p_{k\beta} \p_{k\beta} \right)\p_{i\alpha} V ( \br_i - \br_k ) \delta ( \br - \br_i(t) )
		\\
		& \, = \left( g_\beta \p^2_{\alpha\beta} \Vr - \rho \p^2_{\alpha\beta} ( V* g_\beta) \right) (\br, t) \;,
	\end{aligned}
\end{equation}
where the star $*$ refers to a spatial convolution, {\it e.g.} $(V * g_\beta)(\bx,t) =\int V(\bx-\by) g_\beta(\by,t) d\by$. The dynamic equation for the current density field $\bg$ then follows as
\begin{equation}\label{naka}
	\begin{aligned}
		\tau \p_t g_\alpha + \p_\beta \kappa_{\alpha\beta} &= - g_\alpha - \tau \left( g_\beta \p^2_{\alpha\beta} \Vr - \rho \p^2_{\alpha\beta} \left( V* g_\beta \right) \right)
		\\
		& \quad - \rho \p_\alpha \Vr + \pnt{ 2 D \rho }^{1/2} \Lambda_\alpha \;.
	\end{aligned}
\end{equation}
 The noise term $\Lambda_\alpha$ is Gaussian with correlations
\begin{equation}
	\langle \Lambda_\alpha (\br, t) \Lambda_\beta ( \br', t')\rangle = \delta_{\alpha\beta} \delta (t-t') \delta (\br - \br') \;,
\end{equation}
In Eq.~\eqref{naka} and we have introduced a local tensor $\kappa$ defined by
\begin{equation}\label{stress-t}
	\kappa_{\alpha\beta}  = \tau \sum_{i=1}^N p_{i\alpha} p_{i\beta} \delta ( \br - \br_i (t) ) \;,
\end{equation}
which can be viewed as the fluctuating analog of the kinetic part of the stress tensor (as described by Irving and Kirkwood in their seminal contribution (see Eq.~(5.13) in \cite{Irving-Kirkwood-4:1950}).

As was noted in \cite{Nakamura:09}, with the additional assumption that particles are discernible or that, equivalently for classical particles, there exists a minimal hard-core radius that prevents complete particle overlap unless these are identical, it is possible to rewrite $\kappa$ in a closed form involving $\bg$ and $\rho$. Indeed, if $\delta(\bx-\br_i)\delta(\bx-\br_j)=\delta_{ij} \delta(\bx-\br_i)\delta(\bx-\br_j)$, then
\begin{equation}\label{stress-t2}
	\kappa_{\alpha\beta} (\br, t) = \tau \frac{g_\alpha g_\beta}{\rho}
\end{equation}
In line with the discussion in the introduction of this Section, while the equality in Eq. (\ref{stress-t2}) is obtained working with empirical distributions, it is unclear whether it shall remain valid when passing to the hydrodynamic fields. In any case, the kinetic stress tensor $g_\alpha g_\beta /\rho$ was shown in \cite{Nakamura:09} to be consistent with a kinetic energy $\int \bg^2/(2\rho) d\br$ for equilibrium particles with underdamped dynamics, hence its appeal.


\subsection{A few comments on possible applications}
As discussed in~\cite{Nakamura:09} for equilibrium underdamped
Langevin dynamics, neglecting inertia at the level of the $g$ field is
possible and leads us back to the Dean-Kawasaki
equation~\cite{Dean:96} established directly for equilibrium
overdamped Langevin dynamics. An approximation consisting in
neglecting some of the inertial contributions for AOUPs is the
so-called unified colored-noise one. At the individual particle level,
it amounts to formally equating the lefthand side in
Eq.~\eqref{eq:dyn_active_g} to zero. One could then implement the
Dean-Kawasaki procedure to arrive at a stochastic partial differential
equation for $\rho$ only. The latter would be identical to that
obtained by implementing the procedure outlined in section 3.2
of~\cite{Nakamura:09} that allows one to evaluate the $g$
non-linearity keeping $\rho$ as a slow variable. In a somewhat more
controlled fashion, assuming fast equilibration of the $g$ field in
the small $\tau$ regime, the latter can be enslaved; this procedure,
which basically rests on the replacement
$k_{\alpha\beta}\to \tau D \rho \delta_{\alpha\beta}$, would hold to
first order in $\tau$.

It is, however, much more promising to view Eq.~\eqref{naka}
(complemented with the continuity equation Eq.~\eqref{continuity}) as
a formal basis to generate approximations for dynamical correlations.
To do so, we first rescale the dynamics~\eqref{naka}
and~\eqref{continuity} to work in the small $\tau$ regime. One can
then follow the approach developed for equilibrium dynamics
in~\cite{Das-Mazenko:1986, Dufty:1993} to obtain closed equations for
two-point correlations. To illustrate the starting point of this procedure, we rescale time
and $\bg$ just as we did in Sec.~\ref{sec:fpe}. The resulting
equations read
\begin{equation}\label{eqforg}\begin{split}
\p_t g_\alpha&=-\gamma \left(g_\alpha+\tau \left[g_\beta\p^2_{\alpha\beta}V*\rho-\rho\p^2_{\alpha\beta}(V*g_\beta)\right]\right)\\
&-\rho(\p_\alpha V*\rho)-\p_\beta\left(\frac{g_\alpha g_\beta}{\rho}\right)+\sqrt{2D\rho\gamma}\Lambda_\alpha
\end{split}\end{equation}
where $\gamma=\tau^{-1/2}$, of course supplemented with the local conservation of particles $\p_t\rho+{\boldsymbol{\nabla}}\cdot{\bg}=0$. Written in terms of the local velocity field $\bu=\frac{\bg}{\rho}$ we have fully equivalently that, along with $\p_t\rho=-\boldsymbol{\nabla}\cdot(\rho\bu)$,
\begin{equation}\begin{split}
\p_t \bu+&(\bu\cdot\boldsymbol{\nabla})\bu=-\boldsymbol{\nabla}V*\rho\\
&-\gamma \bu+\sqrt{\frac{2\gamma D}{\rho}}\boldsymbol{\Lambda}\\
&-\gamma\tau \left[(\bu\cdot\boldsymbol{\nabla})(\boldsymbol{\nabla}V)*\rho-\rho\boldsymbol{\nabla}\cdot((\boldsymbol{\nabla}V)*\cdot\bu)\right]
\end{split}
\end{equation}
In an equilibrium framework, the combination $-\gamma\bg+\sqrt{2\gamma
  D\rho}\boldsymbol{\Lambda}$ in Eq.~\eqref{eqforg} expresses exchanges with a thermostat
with an effective friction $\gamma\rho$. The active contribution in
Eq.~\eqref{eqforg}, namely $-\gamma \tau
\left[g_\beta\p^2_{\alpha\beta}V*\rho-\rho\p^2_{\alpha\beta}(V*g_\beta)\right]$,
expresses that friction is enhanced when a minimum of the potential is
reached (which effectively pins particles at such a location and which
entails added effective attraction). Within a fluctuating
hydrodynamics approach the microscopic friction term $\gamma\rho$ is
replaced by a viscosity tensor~\cite{Das-Mazenko:1986}.  We view
Eq.~\eqref{eqforg} as an interesting starting point for controlled
coarse-graining procedures which are deferred for future work.

\section{Conclusion}

Active Ornstein-Uhlenbeck particles were primarily introduced in the
field of active matter for the analytical simplification they offer by
relaxing the non-normality of the active noise. Since fewer studies
exist on AOUPs than on their non-Gaussian counterparts, namely ABPs
and RTPs, their phenomenology has been less thoroughly
investigated. This article brings our knowledge on AOUPs up to par
with that on ABPs and RTPs, as examplified by section~\ref{sec:MIPS}
which report MIPS not only for pairwise forces but also for a new
extension of AOUPs which features quorum-sensing interactions.

Furthermore, a clear gain obtained by working with AOUPs is the
possibility to develop a formal small $\tau$ approximation of the
steady state~\cite{Fodor2016PRL}, which is not limited to the first
order in $\tau$~\cite{Hanggi:87, Cao:93,Fox:86a,
  Fox:86b,Maggi:15,marconi2016velocity}. As shown in
section~\ref{sec:fpe}, the series can be used to obtain
\textit{quantitative} predictions on the steady-state distribution,
and is not limited to capturing qualitative features, as initially
feared by considering only the first order in
$\tau$~\cite{rein2016applicability}.

As shown in section~\ref{sec:TRS}, the small-$\tau$ expansion can be
put to work to characterize the departure of AOUPs from their $\tau=0$
equilibrium limit. In addition, section~\ref{sec:fdr} shows
how linear response can be developed in this non-equilibrium regime,
allowing us to predict the response of the system to a perturbation of
its self-propulsion as well as to external forcings.

Finally, a natural next stage is to build hydrodynamic descriptions of
AOUPs to study both their collective features as well as their
transport properties. The technical tools to do so are mature, and
presented in section~\ref{sec:coll_modes}. We leave these for future
work. For instance, they could potentially help analyze further the
dynamical phase transitions reported recently in models of
self-propelled particles~\cite{Cagnetta2017,Tociu2019,Nemoto2019,Fodor2020,Grandpre2020}

\acknowledgments

\'EF acknowledges support from an ATTRACT Investigator Grant of the
Luxembourg National Research Fund, an Oppenheimer Research Fellowship
from the University of Cambridge, and a Junior Research Fellowship
from St Catharine's College. CN acknowledges the support of Aide
Investissements d’Avenir du LabEx PALM (ANR-10-LABX-0039-PALM). JT and
DM acknowledge support from the ANR grant Bactterns. Work funded in
part by the European Research Council under the Horizon 2020
Programme, ERC grant agreement number 740269. MEC is funded by the
Royal Society.


\appendix

\section{Stationary Measure for one particle in a one-dimensional domain}\label{app:SM}
At order $\tau^{2}$, the exact one-dimensional stationary measure
takes the form
\begin{widetext}
\begin{align}
&\nonumber {P}_{\rm s}(r,p)= e^{-\frac{\phi}{D}-\frac{p^{2}}{2D}}\bigg{[]}c+\tau \left(\frac{3c\phi^{(2)}}{2}-\frac{c\phi^{(1)\ 2}}{2D}-c\frac{p^{2}\phi^{(2)}}{2D}+c_{1}\right)+\tau^{\frac{3}{2}}\left(-\frac{1}{2}cp\phi^{(3)}+\frac{cp^{3}\phi^{(3)}}{6D}\right)\\ \nonumber
&+\tau^{2}\bigg{(}-p^{4}\frac{c\phi^{(4)}}{24D}+p^{4}\frac{c\phi^{(2)\ 2}}{8D^{2}}+p^{2}\frac{c\phi^{(1)}\phi^{(3)}}{4D} +p^{2}\frac{c\phi^{(1)\ 2}\phi^{(2)}}{4D^{2}}-p^{2}\frac{3c\phi^{(2)\ 2}}{4D}-p^{2}\frac{c_{1}\phi^{(2)}}{2D}+a_{2}\int^r e^{\frac{\phi}{D}}dx \\
& \nonumber-\frac{c_{1}\phi^{(1)\ 2}}{2D}+\frac{c\int\phi^{(1)\ 2}\phi^{(3)}dx}{2D}+c_{2}+\frac{c\phi^{(1)\ 4}}{8D^{2}}+\frac{3c_{1}\phi^{(2)}}{2} +\frac{5c\phi^{(2)\ 2}}{8}-\frac{5c\phi^{(1)}\phi^{(3)}}{4}+\frac{5Dc\phi^{(4)}}{8}\bigg{)}\\
 \nonumber &+\tau^{\frac{5}{2}}\bigg{(}-a_{2}e^{\frac{\phi}{D}}p-p\frac{c_{1}\phi^{(3)}}{2}+p\frac{c\phi^{(1)\ 2}\phi^{(3)}}{4D}+p\frac{c\phi^{(2)}\phi^{(3)}}{2} +p\frac{5c\phi^{(1)}\phi^{(4)}}{12}-p\frac{7Dc\phi^{(5)}}{24}+p^{3}\frac{c_{1}\phi^{(3)}}{6D}-p^{3}\frac{c\phi^{(1)\ 2}\phi^{(3)}}{12D^{2}} \\
 \nonumber &+p^{3}\frac{c\phi^{(2)}\phi^{(3)}}{4D}-p^{3}\frac{5c\phi^{(1)}\phi^{(4)}}{36D}+p^{5}\frac{c\phi^{(5)}}{120D}-p^{5}\frac{c\phi^{(3)}\phi^{(2)}}{12D^{2}}\bigg{)}\bigg{]}
\end{align}
\end{widetext}
where explicit dependencies on $r$ and $p$ have been omitted when
unambiguous and $c,c_{1},c_{2},a_{2}$ are integration constants. We
can thus compute the constant steady-state current
\begin{align}
  J=-\sqrt{2\pi } a_{2}D^{2} \tau^2 + o(\tau^2)\;.
\end{align}
Integrating over the $p$ variable, we then obtain
\begin{align}\label{app:eq:SM-general}
&P_{\sS}(r)=\sqrt{2\pi D^{2}}e^{-\frac{\phi}{D}}\bigg{[}(c+\tau c_{1})+\tau \left(c\phi^{(2)}-c\frac{\phi^{(1)\ 2}}{2D}\right)\nonumber \\ \nonumber &
+\tau^{2}\bigg{(}\frac{cD\phi^{(4)}}{2}+\frac{c\phi^{(2)\ 2}}{4}-c\phi^{(1)}\phi^{(3)}+c\frac{\int \phi^{(1)\ 2}\phi^{(3)}dx}{2D}\\ &
+c\frac{\phi^{(1)\ 2}\phi^{(2)}}{4D}+c_{1}\phi^{(2)}+a_{2}\int e^{\frac{\phi}{D}}dx\bigg{)}\bigg{]}\,.
\end{align}

The integration constants then depends on the choice of boundary
conditions. First, we consider the case of a particle in an infinite
domain and thus require the potential to be confining. In this case,
the term $e^{-\frac{\phi(r)}{D}}a_{2}\int^r e^{\frac{\phi}{D}}dx$ does
not vanish automatically in the limit $r\rightarrow \pm \infty$ and $a_{2}$ thus has
to vanish. As expected, this leads to a vanishing current $J$. The
constants $c_{i}$ are then set by normalization at each order
$\tau^{i}$. The expression~\eqref{eq:OnePartOneD} in the main text
then comes from exponentiating Eq.~\eqref{app:eq:SM-general}.

Second, in the case of a periodic potential with finite support,
$a_{2}$ must instead be set by requiring the periodicity of the
steady-state distribution (\ref{app:eq:SM-general}). For a system of
size $L$, this gives :
\begin{align}
a_{2}=-\frac{c\int_{0}^{L}\phi^{(1)\ 2}\phi^{(3)}dx}{2D\int_{0}^{L}e^{\frac{\phi}{D}}dx}\,.
\end{align}
Altogether, we arrive at
\begin{widetext}
\begin{equation}
\begin{aligned}
  & P_\sS ( r )\sim  \exp\left[-\frac{\Phi}{D}+\tau\left(\Phi''(r)-\frac{\Phi'(r)^2}{2D}\right)+\tau^{2}\left(\frac{D\Phi^{(4)}(r)}{2} +\frac{\int^{r} \Phi'(y)^2 \Phi^{(3)}(y) dy}{2D}-\Phi'(r)\Phi^{(3)}(r)-\frac{\Phi''(r)^{2}}{4}\right.\right.\\
\nonumber & -\frac{\Phi'(r)^4}{8 D^2}+\frac{3 \Phi'(r)^2 \Phi''(r)}{4D} \left. \left. -\frac{\int_0^r e^{\frac{\phi}D} \int_0^L \Phi^{\prime 2}\Phi^{(3)}}{2 D \int_0^L e^{\frac{\phi}D}} + \frac{\Phi'(r)^2\int_0^L e^{-\frac{\phi}D} (\frac{\Phi^{\prime 2}}{2 D}-\Phi^{(2)}) }{2 D \int_0^L e^{-\frac{\phi}D}}\right)+{\cal O}(\tau^{3})\right]\,.
\end{aligned}
\end{equation}
\end{widetext}
The last lines differs from the case of a confining potential; it
enforces periodic boundary conditions and leads to the non-vanishing
current~(\ref{eq:CurrentSmooth}).

\section{Approximate dynamics}\label{app:AOUP}

We present in this Appendix two approximate forms of the dynamics of interacting persistent self-propelled particles. They are inspired by approximation schemes which were originally proposed for non-interacting particles: the unified colored-noise approximation~\cite{Hanggi:87, Cao:93}, and the Fox theory~\cite{Fox:86a, Fox:86b}.

We derive the stationary distribution of the unified colored-noise approximation (UCNA). To this aim, we turn the multiplicative Langevin equation, written with the Stratonovich convention in~\eqref{eq:dyn_ucna}, into an additive one by introducing the following change of variables:
\begin{equation}
	q_{i\alpha} = \int \M_{i\alpha, j\beta} \dd r_{j\beta} ,
\end{equation}
so that the Jacobian between $\bq_i$ and $\br_i$ is given by $ \abs{ \det \M } $. The corresponding Langevin equation reads
\begin{equation}
	\dot q_{i\alpha} = - \f{ \p \Phi }{ \p r_{i\alpha} } - (2 D)^{1/2} \eta_{i\alpha} .
\end{equation}
One can explicitly check that this is an equilibrium Langevin dynamics
for $\bq$ because
\begin{equation}
  \f{ \p \Phi }{ \p r_{i\alpha} }= \f{ \p [\Phi+\frac\tau 2 (\nabla_\br \Phi)^2]}{ \p q_{i\alpha} }
\end{equation}
which means that the force in $\bq$-space, and thus in $\br$-space, is
conservative. It is then straightforward to write the corresponding
Fokker-Planck equation, from which we deduce the stationary
distribution as
\begin{equation}
	P_\sS (\cur{ \bq_i } ) \sim \exp \pnt{ - \f{1}{D} \int \f{ \p \Phi }{ \p r_{i\alpha} } \dd q_{i\alpha} } ,
\end{equation}
leading to
\begin{equation}
	P_\sS (\cur{ \br_i } ) \sim \exp \pnt{ - \f{1}{D} \int \f{ \p \Phi }{ \p r_{i\alpha} } \M_{i\alpha, j\beta} \dd r_{j\beta} } \abs{ \det \M } .
\end{equation}
Using $ \M_{i\alpha, j\beta} = \delta_{ij} \delta_{\alpha\beta} + \tau
\p^2_{i\alpha, j\beta} \Phi $, the explicit expression of the
stationary distribution~\eqref{eq:Ps_ucna} follows directly. Note that
the above derivation shows that UCNA does not solely provide an
effective potential for the dynamics: it is a purely equilibrium
approximation of the latter, and hence it is unable to capture
non-equilibrium features such as the emergence of currents in ratchet
potentials as discussed in section~\ref{sec:current}.

Another approximate dynamics can be derived by using functional calculus on the weight $\cP$ of a given time realization of the noises. Introducing the kernel $\cK$ defined in terms of the self-propulsion correlations as
\begin{equation}\label{eq:KK}
	\int \cK (t-u) \avg{ v_{i\alpha} (s) v_{j\beta} (u) } \dd u = \delta (t-s) \delta_{ij} \delta_{\alpha\beta} ,
\end{equation}
we write the probability weight $\cP$ as
\begin{equation}\label{eq:PP}
	\cP = \exp \brt{ - \f{1}{2} \iint_0^t \cK (u-s) v_{i\alpha} (u) v_{i\alpha} (s) \dd u \dd s } .
\end{equation}
The distribution of positions at a given time $t$ can be expressed in terms of this probability weight as
\begin{equation}
	P ( \cur{ \br_i }, t) = \int \cP \prod_{k=1}^N \delta \brt{ \br_k - \bq_k (t) } \cD \bv_k ,
\end{equation}
where the positions $ \bq_k $ satisfy the dynamics $ \dot \bq_k = - \nabla_k \Phi + \bv_k $. It follows that the time derivative of this distribution reads
\begin{equation}
	\begin{aligned}
		\p_t P &= \nabla_i \cdot \Big \{ \int \cP \brt{ \nabla_i \Phi - \bv_i } \prod_{k=1}^N \delta \brt{ \br_k - \bq_k (t) } \cD \bv_k \Big \}
		\\
		&= - \nabla_i \cdot \Big \{ \int \cP \bv_i(t) \prod_{k=1}^N \delta \brt{ \br_k - \bq_k (t) } \cD \bv_k \Big \}
		\\
		& \quad + \nabla_i \cdot \pnt{ P \nabla_i \Phi } .
	\end{aligned}
\end{equation}
To proceed further, we note that
\begin{equation}
	\begin{aligned}
		v_{i\alpha} (t) \cP &= \cP \int \delta (t-s) v_{i\alpha} (s)  \dd s
		\\
		&= - \int_0^\infty \avg{ v_{l\gamma} (t) v_{l\gamma} (s) } \f{ \delta \cP }{ \delta v_{i\alpha} (s) } \dd s ,
	\end{aligned}
\end{equation}
where we have used Eqs.~\eqref{eq:KK} and~\eqref{eq:PP}. It leads to
\begin{equation}
	\begin{aligned}
		& \int \cP v_{i\alpha} \prod_{k=1}^N \delta \brt{ \br_k - \bq_k (t) } \cD \bv_k
		\\
		& \, = - \int_0^\infty \avg{ v_{l\gamma} (s) v_{l\gamma} (t)} \dd s \int \f{ \delta \cP }{ \delta v_{i\alpha} (s) }
		\\
		& \quad  \prod_{k=1}^N \delta \brt{ \br_k- \bq_k (t) } \cD \bv_k
		\\
		\label{eq:IBP}
		& \, = - \nabla_j \cdot \int_0^\infty \avg{ v_{l\gamma} (s) v_{l\gamma} (t) } \dd s \int \cP \f{ \delta \bq_j (t) }{ \delta v_{i\alpha} (s) }
		\\
		& \quad  \prod_{k=1}^N \delta \brt{ \br_k - \bq_k (t) } \cD \bv_k ,
	\end{aligned}
\end{equation}
where we have integrated by parts with respect to $v_{i\alpha}$ to get the second line. From the dynamics $ \dot \bq_i = - \nabla_i \Phi + \bv_i $, we obtain the following identity
\begin{equation}\label{eq:ode}
	\f{\dd}{\dd t} \f{ \delta q_{j\beta} (t) }{ \delta v_{i\alpha} (s) } = - \f{ \delta q_{j\beta} (t) }{ \delta v_{i\alpha} (s) } \frac{\partial^2\Phi}{\partial q_{i\alpha}\partial q_{j\beta}} +  \delta_{ij} \delta_{\alpha\beta} \delta(t-s) .
\end{equation}
This equation contains a sum which was omitted in~\cite{Brader:15}. Introducing the Hessian $\mH$ with elements $ \mH_{i\alpha, j\beta} = \p^2 \Phi / (\p q_{i\alpha} \p q_{j\beta}) $, the solution can be written for $ t > s $ as
\begin{equation}
	\f{ \delta q_{j\beta} (t) }{ \delta v_{i\alpha} (s) } = \brt{ \ee^{ - \int_s^t \mH (w) \dd w } }_{i\alpha, j\beta} .
\end{equation}
Substituting in Eq.~\eqref{eq:IBP} and using $ \avg{ v_{l\gamma} (t) v_{l\gamma} (s) } = D \ee^{ - \abs{ t-s } } / \tau $, we get
\begin{equation}
	\begin{aligned}
		&\int \cP v_{i\alpha} \prod_{k=1}^N \delta \brt{ \br_k - \bq_k (t) } \cD \bv_k
		\\
		& \, = \f{D}{\tau} \p_{j\beta} \Big \{ \int \cP \int_0^t \ee^{ - (t-s) / \tau } \brt{ \ee^{ - \int_s^t \mH (w) \dd w } }_{i\alpha, j\beta} \dd s
		\\
		& \quad  \prod_{k=1}^N \delta \brt{ \br_k - \bq_k (t) } \cD \bv_k \Big \}
		\\
		&= D \p_{j\beta} \pnt{ P \mD_{i\alpha, j\beta} } ,
	\end{aligned}
\end{equation}
where we have introduced the diffusion tensor $\mD$ as
\begin{equation}
	\begin{aligned}
		\mD (t) &= \int_0^t \ee^{ - (t-s) / \tau } \ee^{ - \int_s^t \mH (w) \dd w } \dd s
		\\
		&= \int_0^t \ee^{ -u / \tau } \ee^{ - \int_{t-u}^t \mH (w) \dd w } \dd u .
	\end{aligned}
\end{equation}
This is valid for any value of $\tau$ since we have not used any approximation at this stage of the derivation. To get rid of the kernel in $\mD$, we assume that $\mH$ varies slowly in time, in the same spirit as the original Fox theory~\cite{Fox:86a,Fox:86b}:
\begin{equation}
	\int_{t-u}^t \mH (w) \dd w \simeq u \mH (t) ,
\end{equation}
yielding
\begin{equation}
	\mD (t) \simeq \int_0^t \ee^{ - u / \tau } \ee^{ - u \mH (t) } \dd u = \int_0^t \ee^{ - u \M(t) / \tau } \dd u ,
\end{equation}
where we have used $ \M_{i\alpha, j\beta} = \delta_{ij} \delta_{\alpha\beta} + \tau \mH_{i\alpha, j\beta} $. Integrating and performing the same approximation as Fox gives
\begin{equation}
	\mD(t) \simeq \tau \Mm (t) \brt{ 1 - \ee^{- t \M (t) / \tau} } \simeq \tau \Mm (t) .
\end{equation}
The Fokker-Planck equation~\eqref{eq:fpe_fox} follows directly.


\section{Time reversal, time-translational invariance and time-reversed process}
\label{sec:TTIandTRS}

To discuss time-reversal symmetry, we introduce the Markov process
that corresponds to the time-reversed of the original one. The latter
is defined as follows: it is the Markov process that generates a
time-reversed trajectory ${\br(t_f-t),-\bp(t_f-t)}$ with the same
probability as the original process generates the original trajectory
${\br(t),\bp(t)}$, where $t_f$ is the length of the trajectory. We
denote $\mathcal{L}^{\dagger}$ and $\mathcal{L}_r^{\dagger}$ the
generators of the original Markov process and of its time-reversed
counterpart, respectively. Note that the reversed process is not
obtained by simply reversing time in the original dynamics; the
dynamics obtained in this way would not satisfy the Markov
property. This is well known in the mathematical literature; the
reversed process was rigorously constructed by Haussmann and Padoux in
\cite{haussmann1986time}, see~\cite{norris1998markov} for a detailed
presentation.

Introducing the operator $\Pi f(\bq,\bp)=f(\bq,-\bp)$, time-translation
invariance imposes that
\begin{equation}\label{def:TTIapp}
\langle B(t) A(0) \rangle = \langle B(0) A(-t) \rangle = \langle \Pi B(0) \Pi A(t) \rangle_r
\end{equation}
where the last equality can be regarded as a mathematical
definition of the reversed process. The last average
in~\eqref{def:TTIapp} is then defined as
\begin{equation}\label{eq:defaveragereverse}
  \langle \Pi B(0)\Pi A(t)\rangle_r \equiv \int d \bq d\bp P^\R_s(\bq,\bp) B(\bq,-\bp) e^{t
    \cL_r^\dagger(\bq,\bp)} A(\bq,-\bp)\;.
\end{equation}
where $P_s^\R$ is the stationnary distribution of the time-reversed
process, which satistfies: $P_s^\R(\bq,\bp)=\Pi P_s(\bq,\bp)$. From
Eq.~\eqref{def:TTIapp} we thus have
\begin{equation}
	\begin{aligned}
& \int  d\bq d\bp \,P_{s} (\bq,\bp) A(\bq,\bp) \left(e^{t\mathcal{L}^{\dagger}(\bq,\bp)} B\right)(\bq,\bp) =\\
&\int d \bq d\bp (\Pi P_s)(\bq,\bp) (\Pi B)(\bq,\bp) e^{t
    \cL_r^\dagger(\bq,\bp)} (\Pi A)(\bq,\bp)
	\end{aligned}
\end{equation}
This equality holds for any time $t$ and observables $A$, $B$.  Taking
its time derivative and setting $t=0$ leads, after integrating
by parts, to
\begin{equation}\label{eq:defreversed}
  \cL_r^\dagger=\Pi P_s^{-1} \cL P_s \Pi
\end{equation}
Note that Eq.~\eqref{eq:defreversed} also implies that
\begin{equation}\label{eq:reverseexp}
  e^{t \cL_r^\dagger}=\Pi P_s^{-1} e^{t \cL} P_s \Pi\;.
\end{equation}
A check that $\cL_r^\dagger$ defined as~\eqref{eq:defreversed} indeed
satisfies Eq.~\eqref{def:TTIapp} then stems from rewriting the
right-hand side of Eq.~\eqref{eq:defaveragereverse} as
\begin{equation}
  \begin{aligned}
  \langle \Pi B(0)\Pi A(t)\rangle_r=& \int d \bq d\bp [\Pi P_s(\bq,\bp) \Pi B(\bq,\bp)]\\ &\Pi P_s^{-1} e^{t \cL} P_s \Pi\, \Pi A(\bq,\bp)\;.
  \end{aligned}
\end{equation}
which leads to
\begin{equation}
  \langle \Pi B(0)\Pi A(t)\rangle_r= \int d \bq d\bp B(\bq,\bp)  e^{t \cL} P_s  A(\bq,\bp)\;,
\end{equation}
or, equivalently, to
\begin{equation}\label{eq:TTI}
  \langle \Pi B(0)\Pi A(t)\rangle_r= \langle  B(t) A(0) \rangle\;.
\end{equation}

The definition of time-reversal symmetry is that $\cL_r=\cL$. One then
has that $P_s^\R=P_s$. The left-hand side of~\eqref{eq:TTI} can then be
written as $\langle \Pi B(0)\Pi A(t)\rangle_r=\langle \Pi B(0)\Pi
A(t)\rangle$ so that we finally have
\begin{equation}
  \langle \Pi B(0)\Pi A(t)\rangle= \langle  B(t) A(0) \rangle
\end{equation}
which is Eq.~\eqref{eq:symm-AB-BA-eq-def} of the main text.


\section{Symmetry of time correlations}\label{app:time-correlations}
We present a derivation of the asymmetry of time correlations,
quantified by $\Delta_{AB}$ in~(\ref{eq:Delta-AB})
and~(\ref{eq:Delta-AB-explicit}), using the reverse process introduced
in Appendix~\ref{sec:TTIandTRS}.  When detailed balance holds,
$\mathcal{L}^{\dagger}_r=\mathcal{L}^{\dagger}$ so that the reverse
process is the original dynamics, as expected since equilibrium
dynamics are invariant under time reversal in the steady state. For
AOUPs, using the approximate stationary measure~(\ref{eq:Pss}) and the
definition of $\cL_r^\dagger$ in Eq.~\eqref{eq:defreversed}, we obtain
\begin{equation}\label{eq:app-LHP-AOUPS}
\mathcal{L}^{\dagger}_r(\br,\bar{\bp}) \, = \mathcal{L}^{\dagger}(\br,\bar{\bp}) \, + \tau \mathcal{L}_A^\dagger(\br,\bar{\bp})  + \mathcal{O}(\tau^{3/2})\,
\end{equation}
where $\bar {\bp}=\sqrt{\tau} \bp$ has been introduced in the main
text and the leading-order anti-symmetric part is given by
\begin{equation}\label{eq:app-LHP-AOUPS-Antysymm}
\mathcal{L}^{\dagger}_A (\br,\bar{\bp})=
\Big\{(\nabla_i\nabla^2\Phi)
- \, [(\bar{\bp}_j\cdot \nabla_j)^2\nabla_i\Phi]\Big\}\cdot
\frac{\partial}{\partial \bar{\bp}_i}\,.
\end{equation}
We thus see that detailed-balance holds perturbatively in
$\tau$ in the full $(\br,\bar\bp)$ space, though it only holds up to
order $\sqrt\tau$ when $\bar\bp$ is not integrated out.

We now want to show that the asymmetry of time-correlations
$\Delta_{A,B}$ satisfies the
expansion~(\ref{eq:Delta-AB-explicit}). By definition, we have
\begin{equation}
\langle \Pi A(t) \Pi B(0) \rangle = \int d \bar\bp d\bq P_s(\bq,\bar\bp) (\Pi B)(\bq,\bar\bp) e^{t\cL^\dagger} \Pi A(\bq,\bar\bp)
\end{equation}
Taking the adjoint of Eq.~\eqref{eq:reverseexp} and reorganizing the integral leads to
\begin{equation}
  \langle \Pi A(t) \Pi B(0) \rangle = \int d \bar\bp d\bq (\Pi P_s)(\bq,\bar\bp) A(\bq,\bar\bp) e^{t\cL_r^\dagger} B(\bq,\bar\bp)
\end{equation}
Using the rescaled time $t=\sqrt{\tau}\, \bar t$, we can thus rewrite
$\Delta_{AB}(\bar{t})$ as
\begin{equation}
	\begin{aligned}
\Delta_{AB}(\bar{t}) =
		& \int d\bq d\bar{\bp} \,P_{s}(\bq,\bar{\bp}) A(\bq,\bar{\bp}) \\
&\left( e^{\bar{t}\mathcal{L}^{\dagger}(\bq,\bar{\bp})}-
\frac{P_{s}(\bq,-\bar{\bp})}{P_{s}(\bq,\bar{\bp})}\,e^{\bar{t}\mathcal{L}^{\dagger}_r(\bq,\bar{\bp})}\right)B(\bq,\bar{\bp})\,.
	\end{aligned}
\end{equation}
Note that, for Kramers dynamics~\cite{pavliotis2016stochastic},
$\mathcal{L}^{\dagger}_r=\mathcal{L}^{\dagger}$ and
$P_{s}(\bq,-\bar{\bp})=P_{s}(\bq,\bar{\bp})$ implies that
$\Delta_{A,B}(\bar{t})=0$, as expected. For AOUPs, using the
stationary measure (\ref{eq:Pss}) to expand
$P_{s}(\bq,-\bar{\bp})/P_{s}(\bq,\bar{\bp})$ for small $\tau$, we find
\begin{equation}\label{eq:app-DeltaAB-int}
	\begin{aligned}
&\Delta_{A,B}(\bar{t})
= \int d\bq\,d\bar\bp\, P_{s}(\bq,\bar{\bp})\, A(\bq,\bar{\bp})\\
&\left(e^{\bar{t}\,\mathcal{L}^{\dagger}(\bq,\bar{\bp})} -
 e^{\bar{t}\mathcal{L}^{\dagger}_r(\bq,\bar{\bp})} \right)B(\bq,\bar{\bp}) +\mathcal{O}(\tau^{3/2})\,.
	\end{aligned}
\end{equation}
Using linear response~\cite{baiesi2014thermal}, we note that the
difference between the two evolution operators can be evaluated as
\begin{equation}\label{eq:app-Dyson}
  \begin{aligned}
    &\left[ e^{\bar{t}\mathcal{L}^{\dagger}(\bq,\bar{\bp})} - e^{\bar{t}\mathcal{L}^{\dagger}_r(\bq,\bar{\bp})} \right]
    = \tau \int_0^{\bar{t}} ds_1\\
    &e^{s_1\mathcal{L}^{\dagger}(\bq,\bar{\bp})}  \mathcal{L}_A^{\dagger}(\bq,\bar{\bp}) e^{(\bar{t}-s_1)\mathcal{L}^{\dagger}(\bq,\bar{\bp})} + O\left(\tau^2\right)\,
  \end{aligned}
\end{equation}
where $\mathcal{L}_A^{\dagger}(\bq,\bar{\bp})$ is the leading order antisymmetric part of the generator of AOUPs dynamics, given in (\ref{eq:app-LHP-AOUPS-Antysymm}). Plugging (\ref{eq:app-Dyson}) in (\ref{eq:app-DeltaAB-int}) we obtain
\begin{equation}\label{eq:app-DeltaAB-fin}
  \begin{aligned}
    &\Delta_{A,B}(\bar{t})
    =\tau
    \int_0^{\bar{t}} ds_1\int d\bq\,d\bar\bp\, P_{s}(\bq,\bar{\bp})\, A(\bq,\bar{\bp})\\
    &e^{s_1\mathcal{L}^{\dagger}(\bq,\bar{\bp})}  \mathcal{L}^{\dagger}_A(\bq,\bar{\bp}) e^{(\bar{t}-s_1)\mathcal{L}^{\dagger}(\bq,\bar{\bp})} B(\bq,\bar{\bp})
    +\mathcal{O}(\tau^{3/2})\,.
  \end{aligned}
\end{equation}
Finally, using the explicit expression of $\mathcal{L}^{\dagger}_A$ and rescaling time as $s_1=u_1\sqrt{\tau}$, we obtain eq. (\ref{eq:Delta-AB-explicit}), which was the goal of this Appendix.

\section{Perturbation of the self-propulsion speed}\label{app:pert_temp}
We present in this Appendix a derivation of Eq. (\ref{eq:entr-flux-AOU-new-time}). To do so, we apply the Agarwal formula
\cite{agarwal1972,Baiesi:13} which states that the response of an
observable $A$ to the perturbation $D\to D(1 + h \Theta(t))$ is given
by
\begin{equation}\label{eq:app:Agarwall}
\langle \delta A(\bar{t})\rangle = \int_0^{\bar{t}} ds \left\langle \frac{\mathcal{L}_p P_{s}}{P_{s}}(0)\, A(s) \right\rangle+{\cal O}(h^2)
\end{equation}
where $\mathcal{L}_p$ is the Fokker-Planck operator corresponding to the perturbation
\begin{equation}
  \mathcal{L}_p(\bar{\bp})=
  \frac{h D}{\sqrt{\tau}}\frac{\partial^2}{ \partial \bar{\bp}_i^2}\,.
\end{equation}

In order to obtain Eq. (\ref{eq:entr-flux-AOU-new-time}) from
(\ref{eq:app:Agarwall}), we first note that, inserting
expression~\eqref{eq:Pss} for $P_s$ into~\eqref{eq:app:Agarwall} leads
to
\begin{equation}\label{eq:Argawall-step-1}
\frac{1}{h}\frac{\mathcal{L}_p(\bar{\bp}) P_{s}(\bq,\bar{\bp})}{P_{s}(\bq,\bar{\bp})} = - \frac{1}{D} \mathcal{L}_r^{\dagger}(\bq,-\bar{\bp})\,H_{\rm eff}(\bq,\bar{\bp}) +O(\tau)
\end{equation}
where $H_{\rm eff}$ is defined as:
\begin{equation}
  \begin{aligned}
    H_{\rm  eff}&=H_0+\tau  H_1 
    \\
    H_0&=\Phi + \frac{\bar{\bp}_i^2}{2}\\
    H_1&=  \frac{1}{2}\left[
      \pnt{ \nabla_i \Phi }^2 + \pnt{ \bar{\bp}_i \cdot \nabla_i }^2 \Phi - 3 D \nabla_i^2 \Phi
      \right] \,.
  \end{aligned}
\end{equation}
Eq.~\eqref{eq:app:Agarwall} can thus be rewritten as
\begin{equation}\label{eq:app:Agarwall2}
\langle \delta A(\bar{t})\rangle = -\frac h D \int_0^{\bar{t}} ds \left\langle (\Pi \cL_r^\dagger) H_{\rm eff}(0)\, A(s) \right\rangle+{\cal O}(\tau,h^2)
\end{equation}
Then, from the definition of time-translational invariance, we have that
\begin{equation}
  \begin{aligned}
    & \int  d\bq d\bp \,P_{s} (\bq,\bp) H_{\rm eff}(\bq,\bp) \left(e^{s\mathcal{L}^{\dagger}(\br)} A\right)(\bq,\bp) =\\
    &\int d \bq d\bp (\Pi P_s)(\bq,\bp) (\Pi A)(\bq,\bp) e^{s
      \cL_r^\dagger(\bq,\bp)} (\Pi H_{\rm eff})(\bq,\bp)
  \end{aligned}
\end{equation}
Taking the time derivative of this equality at time $s=0$ and using
Eq.~\eqref{eq:reverseexp} then lead to
\begin{equation}
  \langle H(0) L^\dagger A(s) \rangle = \langle A(s) (\Pi L_r^\dagger) H_{\rm eff}(0) \rangle
\end{equation}
Using that $\partial_s A(s)=L^\dagger A(s)$,
Eq.~\eqref{eq:app:Agarwall2} can thus be finally rewritten as
\begin{equation}\label{eq:app:Agarwall3}
  \langle \delta A(\bar{t})\rangle = -\frac h D  \int_0^{\bar{t}} ds \left\langle H_{\rm eff}(0)\, \frac{ d A(s)}{ds} \right\rangle+{\cal O}(\tau,h^2)
\end{equation}
Then, using $\bar{t}=s\sqrt{\tau}$ with $s\sim\mathcal{O}(\tau^0)$ and integrating, we obtain Eq.~(\ref{eq:entr-flux-AOU-new-time}).

Note that, again, we did not use the path-integral techniques
described in Section~\ref{subsec:perturbing-potential} for describing
perturbations obtained by applying an external force. The operatorial
approach we followed here holds for short trajectories, albeit we
expect the result to be valid for longer ones. An interesting
alternative approach would be to generalize the analysis presented
in~\cite{Falasco:16, Falasco2016} which uses path integral formulation
to solve a similar problem, albeit in equilibrium.



\bibliographystyle{apsrev4-1}
\bibliography{Biblio}

\end{document}